\documentclass[10pt, conference]{IEEEtran}

\IEEEoverridecommandlockouts

\usepackage{cite}
\usepackage{balance}

\usepackage{amsmath,amssymb,amsfonts}
\usepackage{graphicx}
\usepackage{textcomp}
\usepackage{xcolor}
\usepackage{threeparttable}

\usepackage{xspace}
\usepackage{algorithmicx}
\usepackage{algorithm}
\usepackage[noend]{algpseudocode}
\usepackage{tabularx}
\usepackage{multirow}
\usepackage{booktabs}
\usepackage{makecell}
\usepackage[figuresleft]{rotating}
\usepackage{tikz}
\usepackage{pdflscape}

\usepackage{hyperref} 
\hypersetup{colorlinks = true,
	    linkcolor = black, 
	    urlcolor = blue,
            citecolor = red}

\newcommand{\circled}[1]{%
    \tikz[baseline=(char.base)]{
        \node[shape=circle,draw,inner sep=0.5pt,fill=black,text=white,minimum size=1em] (char) {#1};
    }%
}         
            
\algnewcommand\algorithmicinput{\textbf{Input:}}
\algnewcommand\Input{\item[\algorithmicinput]}
\algnewcommand\algorithmicoutput{\textbf{Output:}}
\algnewcommand\Output{\item[\algorithmicoutput]}

\author{\IEEEauthorblockN{Anonymous Authors}}

\def\BibTeX{{\rm B\kern-.05em{\sc i\kern-.025em b}\kern-.08em
    T\kern-.1667em\lower.7ex\hbox{E}\kern-.125emX}}

\newcommand{\SysName}{\texttt{AMSP}\xspace}

\begin{document}

\title{\SysName: Reducing Communication Overhead of ZeRO for Efficient LLM Training}

\author{
    \IEEEauthorblockN{
    Qiaoling\ Chen\IEEEauthorrefmark{1}\IEEEauthorrefmark{2}\IEEEauthorrefmark{3},
    Qinghao Hu \IEEEauthorrefmark{1}\IEEEauthorrefmark{2}\IEEEauthorrefmark{3},
    Guoteng Wang \IEEEauthorrefmark{3},
    Yingtong Xiong \IEEEauthorrefmark{3},
    Ting Huang \IEEEauthorrefmark{4}, 
    Xun Chen \IEEEauthorrefmark{4} \\ 
    Yang Gao \IEEEauthorrefmark{3},
    Hang Yan \IEEEauthorrefmark{3}, 
    Yonggang Wen \IEEEauthorrefmark{2},
    Tianwei Zhang \IEEEauthorrefmark{2},
    Peng\ Sun \IEEEauthorrefmark{3}\IEEEauthorrefmark{4}
    }\\
    \IEEEauthorblockA{
    \IEEEauthorrefmark{1} S-Lab, NTU, 
    \IEEEauthorrefmark{2} Nanyang Technological University, 
    \IEEEauthorrefmark{3} Shanghai AI Laboratory
    \IEEEauthorrefmark{4} SenseTime}
}

\maketitle
\thispagestyle{plain}
\pagestyle{plain}

\begin{abstract}

Training large language models (LLMs) encounters challenges in GPU memory consumption due to the high memory requirements of model states. The widely used Zero Redundancy Optimizer (ZeRO) addresses this issue through strategic sharding but introduces communication challenges at scale. To tackle this problem, we propose \SysName, a system designed to optimize ZeRO for scalable LLM training. \SysName incorporates three flexible sharding strategies: \emph{Full-Replica}, \emph{Full-Sharding}, and \emph{Partial-Sharding}, and allows each component within the model states (Parameters, Gradients, Optimizer States) to independently choose a sharding strategy as well as the device mesh.
We conduct a thorough analysis of communication costs, formulating an optimization problem to discover the optimal sharding strategy. Additionally, \SysName optimizes distributed LLM training by efficiently overlapping communication with computation. Evaluations demonstrate up to 52\% Model FLOPs Utilization (MFU) when training the LLaMA-based model on 1024 GPUs, resulting in a 1.56 times improvement in training throughput compared to newly proposed systems like MiCS and ZeRO++.

% Training large language models (LLMs) faces challenges in GPU memory consumption due to high model state memory consumption. The widely adopted Zero Redundancy Optimizer (ZeRO) addresses this issue through strategic sharding but introduces communication challenges at scale. To bridge this gap, we propose \SysName, a system optimizing ZeRO for scalable LLM training. \SysName incorporates three flexible sharding strategies—Full-Replica, Full-Sharding, and Partial-Sharding—and allows each component within the model state (Parameter, Gradient, Optimizer State) to independently choose a sharding strategy as well as the device mesh. \SysName further optimizes distributed LLM training by efficiently overlapping communication with computation. We conduct a thorough analysis of memory and communication costs, formulating an optimization problem to discover sharding factors that minimize communication costs. Evaluations show up to 52\% Model FLOPs Utilization when training the LLaMA-based model on 1024 GPUs, which improves the training throughput by 1.56 times compared to newly proposed systems.

\end{abstract}

%-------------------------------------------------------------------------------
\section{Introduction}
\label{sec_intro}

Large Language Models (LLMs) have demonstrated exceptional performance in various tasks, with the relationship between model size and performance often following a power-law relationship. Despite the prevailing trend of training giant models like GPT-3 with 175 billion parameters, recent studies indicate that optimal performance may be achieved with smaller models trained on larger datasets \cite{trainingcomputeoptimal}. Emerging LLMs like LLaMA \cite{LLaMA}, featuring 7 billion to 30 billion parameters.

Training LLMs significantly demands on GPU memory, primarily due to the substantial memory consumption of model states, encompassing parameters ($P$), gradients ($G$), and optimizer states ($OS$). 
Additional memory is allocated for activations and temporary buffers.
For instance, when training LLaMA-7B, a substantial 112GB of memory is required for model states, surpassing the capacity of an 80GB NVIDIA A100 GPU. To address this challenge, ZeRO,  implemented in Deepspeed \cite{DeepSpeed} and PyTorch FSDP \cite{PyTorchFSDP}, introduces a sharding strategy to alleviate redundant memory allocations. ZeRO-1 distributes optimizer states across GPUs, ZeRO-2 further shards gradients and ZeRO-3 extends this approach to parameters, gradients, and optimizer states. This strategic sharding optimizes memory usage, enabling efficient training of large models within GPU constraints. ZeRO could work in cooperation with 3D parallelism \cite{Megatron-LM} and has become widely adopted in distributed LLMs training.

ZeRO heavily relies on collective communication for effective model states management, introducing challenges in large-scale LLM training due to the substantial transmission cost.
In our experiments, training a LLaMA-7B model on 8 GPUs using ZeRO-1 achieves a model FLOPs utilization (MFU) of {$63\%$}, but scaling to 1024 GPUs with the same batch size results in a significant performance reduction, with the MFU dropping to {$36\%$}. 
The costly communication of ZeRO can be attributed to three primary factors: 1) a significant bandwidth discrepancy between inter-node and intra-node networks, 2) an increase in collective communication latency as the communication scale grows, and 3) the use of a small micro-batch size per GPU on numerous GPUs, exacerbating the compute-to-communication ratio imbalance.

Several approaches have been proposed to reduce the communication overhead of ZeRO with improved memory utilization. ZeRO++ \cite{ZeRO++} achieves this by maintaining a secondary parameters shard within small subgroups, effectively reducing communication latency when collecting them.  MiCS \cite{MiCS}  shards all model states components within subgroups and replicates them across subgroups, reducing communication scale and consequently reducing communication latency, leading to enhanced training performance.  
Despite these efforts, when scaling LLM training to a large extent, ZeRO++ and MiCS exhibit suboptimal speedup ratios due to two factors.
Firstly, the inflexible model states sharding mechanism results in suboptimal communication costs. This limitation is evident in the case of MiCS, where scaling LLaMA-7B training from 8 GPUs to 1024 GPUs leads to a significant decrease in model training performance, even falling below the efficiency of ZeRO-1. Secondly, the inefficiency of the overlap mechanism poses a challenge. For instance, an efficient implementation of MiCS with a streamlined communication-computation overlap can outperform DeepSpeed-MiCS by a factor of 2$\times$ during the training of LLaMA-13B on 1024 GPUs. 

% unable to overlap all communications, only consider parts their inflexible model state sharding mechanism.   \todo{say some facts about the overlap problem in existing solutions, also hightlight it.}

We propose \SysName for reducing the communication overhead of ZeRO for training LLMs at scale. To achieve this goal, \SysName incorporates three flexible sharding strategies—\emph{Full-Replica}, \emph{Full-Sharding}, and \emph{Partial-Sharding}—allowing each component within the model states (\emph{P}, \emph{G}, \emph{OS}) to independently choose a sharding strategy. The introduced sharding factors ($s_p^0 \times s_p^1, s_g^0 \times s_g^1, s_{os}^0 \times s_{os}^1$) control the number of GPUs and the device mesh over which the tensors are sharded. Given this flexibility, we analyze the memory consumption and communication costs for each sharding dimension. Then, we formulate an optimization problem aimed at discovering optimal sharding factors that minimize communication costs while adhering to the constraint of GPU memory capacity.  \SysName implements an execution engine tailored for training LLMs, incorporating these flexible sharding factors to achieve optimized communication efficiency during training.

\SysName further optimizes distributed LLM training by efficiently overlapping communication with computation. When parameters sharding is enabled, \SysName employs a strategy to prefetch parameters for the next layer using \texttt{AllGather} during the forward pass, while simultaneously performing current layer computations. In the backward pass, \SysName strategically schedules \texttt{ReduceScatter} operations for gradient synchronization within each parameters sharding subgroup, avoiding conflicts and ensuring that computations continue without waiting for communications to finish. Additionally, with activation re-computation, \SysName carefully manages the additional forward computation in the backward pass, retaining prefetched parameters for immediate use. These overlapping strategies collectively reduce GPU idle time and significantly enhance the training performance of LLMs.

Extensive evaluations show a significant system performance of \SysName on training LLaMA-based models. On 1024 Nvidia Ampere GPUs, the MFU of \SysName is $51\%$, $52\%$, and $42\%$ on LLaMA-7B, LLaMA-13B, and LLaMA-30B training. In comparison, MiCS demonstrates lower MFU values at 35\%, 33\%, and 29\% for the same models, ZeRO++ shows the least MFU among the three, with MFU rates at merely $4\%$, $6\%$, and $5\%$ for the 7B, 13B, and 30B models, respectively. Compared to MiCS and ZeRO++, \SysName improves the training throughput by a factor of $1.4-12.7$ on 1024 GPUs for training LLaMA-based models. 
\SysName \footnote{Please visit \url{https://github.com/InternLM/InternEvo} to access the system.} has been  used for training  InternLM \cite{InternLM} on thousands of GPUs. Our efforts also encompass an exhaustive study characterizing a six-month development workload trace of LLM collected from our GPU datacenter \cite{hu2024characterization}.

% In summary, we make the following contributions:
% % In addition, \SysName successfully copes with the aforementioned deploymenand achieves the following desirable properties:

% \begin{itemize}[]
%       \item 
%       We build a flexible model states partitioning space enabling independent and fine-grained partitioning strategies for $P$,$G$, and $OS$. We systematically analyze the memory and communication costs in this space.
        
%         % the data-dependent and computation-independent properties of the model state to construct a larger partition space than previous DL systems \cite{ZeRO,ZeRO-Infinity,ZeRO++,Megatron-LM,PyTorchFSDP}.

%----------------------------------------------------

\section{Background}
\label{sec_motivation}

We provide a brief introduction to the essential background of  LLM training and the associated challenges to improve performance. Table \ref{tab:notation} gives notations used in this work.

\subsection{LLM Architecture}

LLMs like GPT-3 \cite{GPT3} and LLaMA \cite{LLaMA} widely adopt the Transformer \cite{Attention} architecture with multiple layers. Each Transformer layer comprises a list of modules, such as linear, multi-head-attention (MHA), and norm modules.
The input and output dimensions for each Transformer layer are denoted as $B \times S \times H$, where $B$ represents the micro-batch size, $S$ indicates the sequence length, and $H$ is the hidden dimension.
The relationship between the model size of LLMs and their performance is typically governed by a power-law relationship. While there has been a trend to train giant models like GPT-3 with 175B parameters, existing studies suggest that optimal model performance may be attained with smaller models trained on larger datasets \cite{trainingcomputeoptimal}.  As illustrated in Table \ref{tab:llms_with_parameter_count}, recently introduced LLMs like LLaMA and InternLM typically feature 7B to 30B parameters. 

\begin{table}
    \centering
    \caption{Notations used in this work.}
    \begin{tabular}{@{}ll@{}}
         \toprule
         \textbf{Notation} & \textbf{Meaning} \\
         \midrule 
         $D$ & Memory consumption of a GPU. \\
         $T$ & Time consumption. \\
         $\Phi$ & Model parameters count during training. \\
         ${N}$ & Total number of GPU nodes used for training. \\
         ${R}$ & Number of GPUs per computational node. \\
         ${B}$ & Micro-batch size (sequences per micro-batch). \\
         ${M}$ & Number of micro-batches. \\
         % ${S}$ & Sequence length (tokens in a sequence). \\
         % ${H}$ & Hidden dimension size of the model. \\
         ${L}$ & Number of layers of the model. \\
         ${K}$ & Number of modules within a layer of the model. \\
         $s_{dp},s_{tp},s_{pp}$ & Size of data, tensor and pipeline parallelism. \\
         $s_{p},s_{g},s_{os}$ & Sharding factors of parameters, gradients and model states.\\
         % $\mathbb{C}$ & Combination of $s_{p},s_{g},s_{os}$. \\
         \bottomrule
    \end{tabular}
    \label{tab:notation}
\end{table}

\begin{table}
    \centering
    \caption{Popular LLMs and their parameters count.}
    \begin{tabular}{ll|ll}
        \toprule
        Model        & \# Parameters          & Model   & \# Parameters \\ \midrule
        GPT-3         & 175B                  & BLOOM     & 175B       \\ 
        LLaMA         & 7B, 13B, 33B, 65B     & Mistral   & 7B         \\ 
        LLaMA2        & 7B, 13B, 70B          & InternLM2 & 7B, 20B    \\ 
        Cerebras-GPT  & 1.3B, 2.7B, 6.7B, 13B & Baichuan2 & 7B, 13B    \\ \bottomrule
    \end{tabular}
    \label{tab:llms_with_parameter_count}
\end{table}

\subsection{Distributed LLM Training}

Efficiently training LLMs at scale in GPU clusters involves utilizing  3D parallelism. Data Parallelism (DP) divides input data into chunks, distributing them across GPUs, where each GPU independently computes gradients, later synchronized through  \texttt{AllReduce} communication  \cite{PyTorchDistributed}. Tensor Parallelism (TP)  distributes parameters across GPUs along specific dimensions for parallel training. Megatron-LM employs TP to partition linear layers along the row or column dimension, integrating collective communication operations for consistent results \cite{Megatron-LM}. Pipeline Parallelism (PP) evenly divides a model's Transformer layers into multiple stages, distributing them across GPUs. A scheduler splits an input batch into micro-batches, alternating between forward and backward computations \cite{GPipe} \cite{PipeDream}. Two consecutive pipeline stages exchange intermediate data through point-to-point communication.

\subsection{ZeRO}

\begin{figure}[t]
    \centering
    \includegraphics[width=0.95\linewidth]{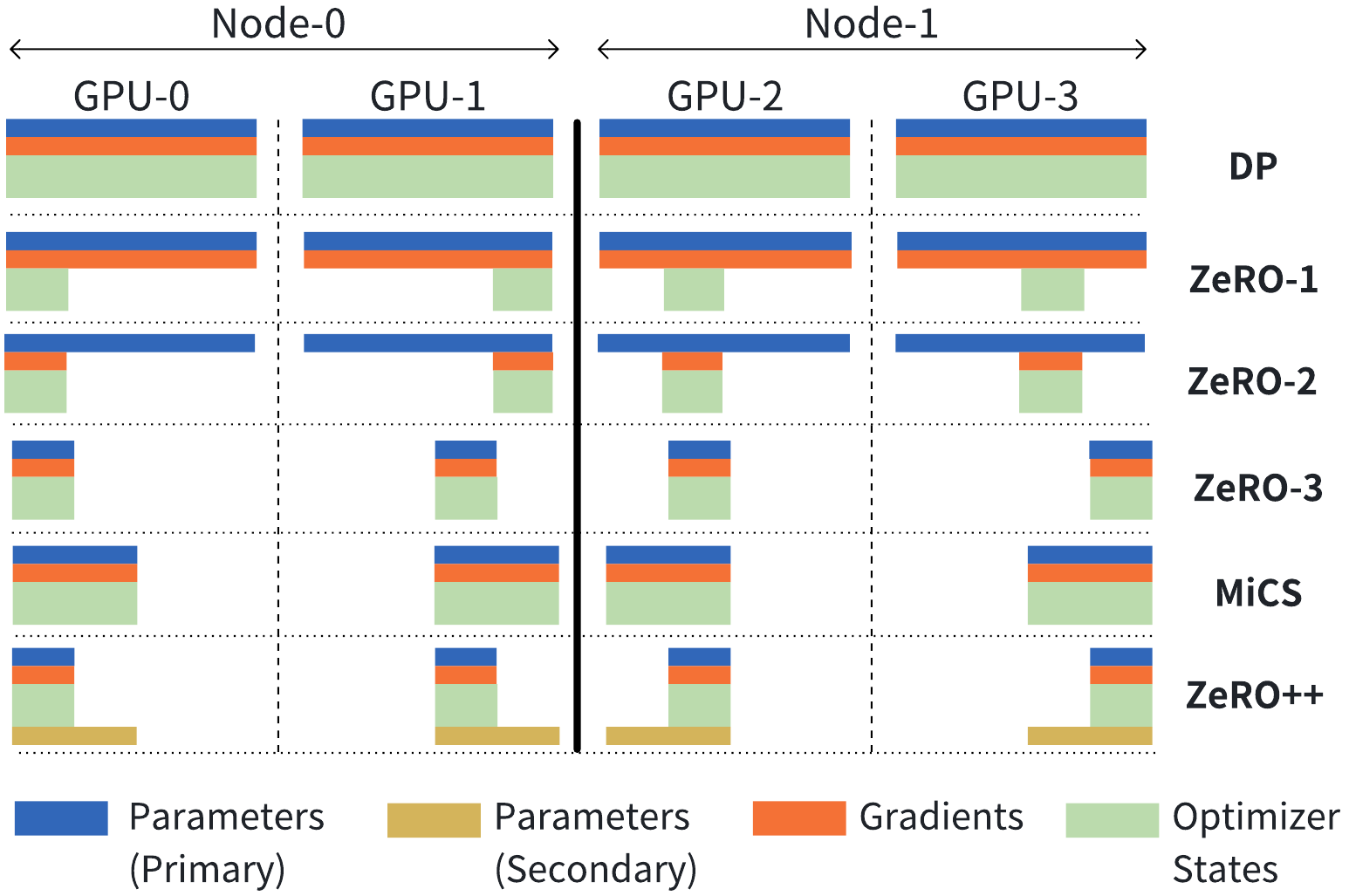}
    \caption{Overview of GPU memory allocation for model states with different strategies. ZeRO-1 and ZeRO-3 significantly reduce memory consumption for model states compared to standard data parallelism. MiCS and ZeRO++ are proposed to mitigate communication overhead, particularly cross-node communication time, in comparison to the ZeRO approach.}
    \label{Fig: zero_overview}
\end{figure}

Training LLMs results in significant memory consumption, largely due to the occupation of GPU memory by model states, which comprise tensors containing parameters (\emph{P}), gradients (\emph{G}), and optimizer states (\emph{OS}). The remaining memory is allocated to activations and temporary buffers. In the context of a model with $\Phi$ parameters, employing mixed precision training alongside the Adam optimizer \cite{Adam}, it necessitates $2\Phi$, $2\Phi$, and $12\Phi$ bytes of GPU memory for \emph{P}, \emph{G} and \emph{OS}, respectively. As an illustrative example, the LLaMA-7B model requires 112GB of memory for its model states, exceeding the memory capacity of an NVIDIA A100 GPU (80GB). As shown in Figure \ref{Fig: zero_overview}, ZeRO reduces redundant memory usage for model training by sharding model states \cite{ZeRO}. 

ZeRO-1 splits optimizer states across GPUs ($s_{os}\!>\!1$). In the training phase, each GPU independently computes gradients through forward and backward computations, which are then synchronized across $s_{dp}$ GPUs using \texttt{AllReduce}. Each GPU updates specific portions of the model parameters. The most recent model parameters for a GPU are gathered from other GPUs using the \texttt{AllGather} operation.  
ZeRO-2 extends this approach by further sharding gradients across GPUs ($s_{g}\!=\!s_{os}\!>\!1$). Each GPU retains only the gradients corresponding to its optimizer states segment after the reduction operation.

ZeRO-3, also implemented in FSDP \cite{PyTorchFSDP}, employs the sharding strategy on model parameters, gradients, and optimizer states ($s_{p}\!=\!s_{g}\!=\!s_{os}\!>\!1$). Before each forward and backward computation, individual GPUs execute the \texttt{AllGather} operation to assemble the complete set of model parameters and subsequently discard them post-computation. The synchronization of gradients across GPUs is achieved through \texttt{Reduce-Scatter}. Each GPU updates its corresponding shard of model parameters using the maintained optimizer states and gradients at the end of each step. 

\section{Challenges and Motivation}

ZeRO has gained extensive adoption across various training frameworks, such as DeepSpeed \cite{DeepSpeed}, FSDP \cite{PyTorchFSDP}, and ColossalAI \cite{li2023colossal}, owing to its user-friendly interface and scalability across hundreds of GPUs. Despite leveraging high-bandwidth RDMA networks, challenges emerge in the form of poor Quality of Service (QoS) during distributed LLM training on large-scale GPU clusters. This can be mainly attributed to significant communication overhead.

\subsection{High Communication Overhead}
\label{subsec:highcommcost}

ZeRO necessitates extensive usage of collective communication for managing parameters and gradients. The transmission cost across large-scale clusters presents a challenge, as it cannot be easily mitigated through computation-communication overlapping.
When training a LLaMA-7B model on 8 GPUs using ZeRO-1, the model FLOPs utilization (MFU) attains $63\%$ in our test-bed. Scaling the training to 1024 GPUs with the same batch size results in a notable performance reduction, with the MFU dropping to $36\%$. Similarly, scaling LLaMA-13B training from 8 GPUs to 1024 GPUs with ZeRO-3 leads to a substantial MFU reduction from $47\%$ to $4\%$.

\begin{figure}[t]
    \centering
    \includegraphics[width=\linewidth]{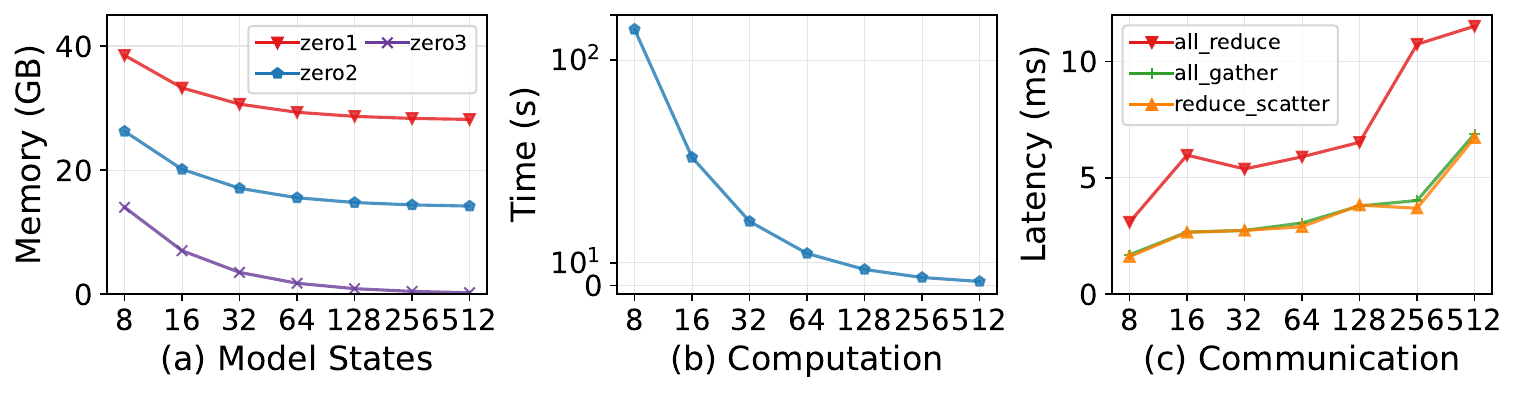}
    \caption{Micro-benchmark of training LLaMA-7B across a scale of GPUs, ranging from 8 to 512, while maintaining a global batch size of 4M tokens. The micro-batch size $B$ is consistently set to $1$ in all tests. Panel (a) illustrates the GPU memory consumption of model states. Panel (b) depicts the time taken for forward and backward computations. Panel (c) presents the latency of three communication operations with a fixed message size of 256MB.}
    \label{fig: motivation}
\end{figure}

\begin{figure}[t]
    \centering
    \includegraphics[width=\linewidth]{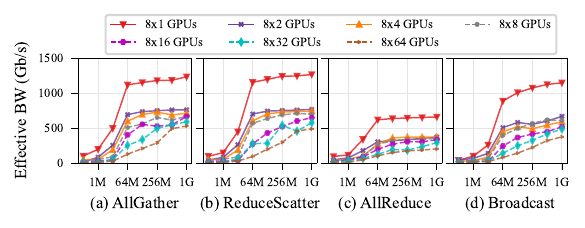}
    \caption{Performance evaluation of collective communication operations using NCCL. The assessment is conducted with varying message sizes (in bytes). GPU nodes are linked using 4 Mellanox Infiniband HDR NICs (800 Gbps bandwidth in total). The notation $8 \! \times \! A$ GPUs indicates that the tests were conducted on $A$ nodes, with each node housing 8 NVIDIA Ampere GPUs (A800) connected by NVLINK.}
\label{fig: commu_bench}
\end{figure}

Three main factors contribute to the costly communications for large-scale LLM training with ZeRO.
Firstly, there exists a notable discrepancy between inter-node network bandwidth and intra-node NVLINK bandwidth. High-performance DGX-A100 nodes offer 600GB/s intra-node bidirectional bandwidth per GPU and provide 400GB/s  inter-node bidirectional bandwidth per node. The bandwidth ratio between intra-node and inter-node measures at 2 in our test-bed.
Secondly, the latency of collective communication operations demonstrates a positive correlation with communication scale  \cite{ByteScheduler}\cite{sun2019gradientflow}\cite{sun2016timed} and illustrated in Figure \ref{fig: motivation}(c). Figure \ref{fig: commu_bench} further illustrates a reduction in the effective bandwidth of communication operations utilized by ZeRO, scaling from 8 GPUs to 512 GPUs. Thirdly, the global batch size limitation for convergence efficiency imposes the use of a very small batch size per GPU when training on numerous GPUs. 
As depicted in \ref{fig: motivation} (b), the computation time of the LLaMA-7B model training linearly reduces from 8 GPUs to 512 GPUs while maintaining a consistent 4M batch size. 
This reduction adversely affects the compute-to-communication ratio,  leading to a communication bottleneck.

\begin{figure*}[t]
    \centering
    \includegraphics[width=\linewidth]{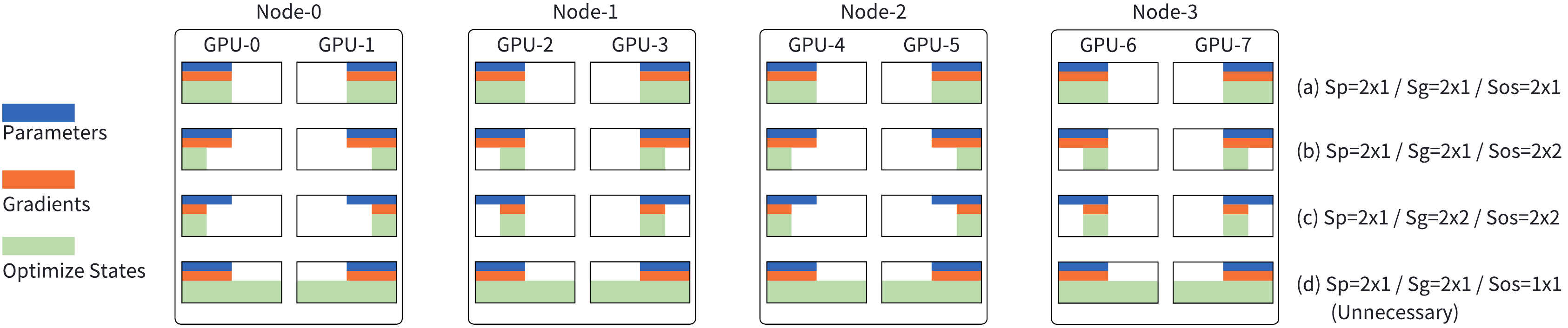}
    \caption{Optimizing model states sharding through the dependency rule. In this instance, when $s_p=s_g=2$, there's no need to set $s_{os}=1$ as it would store redundant optimized states, incurring additional communication costs.}    
    \label{fig: dependencebased}
\end{figure*}

\subsection{Trade-off between Communication and Memory}

A trade-off exists between memory utilization and communication cost in distributed LLM training. 
Initially, the communication cost can be effectively reduced by diminishing the communication scale. This involves limiting communications to a smaller group of GPUs, potentially within the same node, which mitigates the overall communication cost. In addition, as depicted in Figure \ref{fig: motivation} (a), scaling  ZeRO to a large scale does not yield substantial memory savings compared to a smaller size. Consequently, various approaches have been proposed to reduce communication overhead with higher memory usage.

ZeRO++ \cite{ZeRO++} keeps a secondary shard of parameters while sharding other model states across the cluster ($s_p\!=\!s_g\!=\!s_{os}\!=\!s_{dp}$), as shown in Figure \ref{Fig: zero_overview}. In the forward phase, it collects parameters across all GPUs and maintains a secondary shard of parameters within a small subgroup of GPUs, potentially within the same node. During the backward phase, it collects parameters from this secondary shard. Additionally, ZeRO++ uses quantization to compress parameters and gradients, effectively reducing inter-node communication size. Note that we would not enable configurations related to the quantization of ZeRO++ to ensure consistent model quality.

MiCS \cite{MiCS} and FSDP \cite{PyTorchFSDP} facilitate the sharding of model states within a subgroup and replicate them across subgroups ($s_p\!=\!s_g\!=\!s_{os}\!<\!s_{dp}$), as shown in Figure \ref{Fig: zero_overview}. These approaches employ \texttt{AllGather} to collect parameters within a subgroup for both forward and backward computation and synchronize gradients across the cluster using \texttt{ReduceScatter}. Consequently, MiCS and FSDP contribute to improved training performance by effectively reducing the communication scale. It is crucial to configure an appropriate subgroup size to prevent Out-Of-Memory (OOM) errors.

\subsection{Motivation}

Despite efforts to reduce communication costs, ZeRO++ and MiCS still exhibit poor speedup ratios when scaling  LLM training to a large scale. This is attributed to their inflexible model states sharding mechanism,  requiring $s_p\!=\!s_g\!=\!s_{os} \! \leq \! s_{dp}$  in all cases. Such a configuration may not be optimal for training LLMs with diverse model sizes and hyper-parameters. In addition, the inefficiency of the overlap mechanism in ZeRO++ and MiCS also poses a challenge. For instance, when scaling LLaMA-7B training from 8 GPUs to 1024 GPUs with MiCS, MFU decreases from $50\%$ to $35\%$ in our test-bed. In this scenario, MiCS even exhibits lower performance than ZeRO-1, highlighting the drawbacks of the inflexible model states sharding mechanism.

In this study, the three components of model states (\emph{P}, \emph{G}, \emph{OS}) are sharded into independent subgroups and replicated across these subgroups, following the condition $s_p\!\leq\!s_{dp},s_g\!\leq\!s_{dp},s_{os}\!\leq\!s_{dp}$.
This flexibility allows us to fine-tune the trade-off between communication and GPU memory by configuring $s_p, s_g, s_{os}$. By doing so, we may achieve minimal communication cost for distributed LLM training through individualized configuration of the communication scale on \emph{P}, \emph{G}, and \emph{OS}, while respecting GPU memory constraints. 

Taking LLaMA-7B as an illustrative example, we adopt the configuration of $s_p\!=\!s_g\!=\!1, s_{os}\!=\!8$ for training. In this setting, each GPU retains a complete copy of parameters and gradients, while each node stores a full copy of optimization states. During training, gradients are synchronized across clusters using \texttt{AllReduce}, and each GPU obtains the latest parameters within the same node through \texttt{AllGather} at the end of each step. In our test-bed, scaling LLaMA-7B training from 8 GPUs to 1024 GPUs with this configuration results in an acceptable MFU reduction from $64\%$ to $51\%$.

\section{Model States Sharding and Analysis}

In this section, we assume that there is no tensor parallelism or pipeline parallelism during the training, which implies that  \(s_{tp}=s_{pp}=1\). This simplification allows us to focus on the impact of the discussed sharding strategies on communication and memory aspects.

\subsection{Performance Model of Collective Communication}
\label{subsec:CommperformanceModel}

The $\alpha-\beta$ cost model \cite{thakur2003improving} is widely employed to characterize the performance of collective communication \cite{zhang2023dear}. Taking the example of a ring-based \texttt{AllReduce} on $p$ GPUs, where the input size is $v$, and the physical bandwidth between two GPUs is $w$, the input is evenly split into $p$ chunks. In the first stage, each chunk undergoes $p-1$ rounds of reduction to each GPU, constituting a \texttt{ReduceScatter} operation with a time complexity of $t_{rs}=(p-1)(\alpha+\frac{v}{w \times p})$, where $\alpha$ denotes the latency per transmission. Then, each reduced chunk at every GPU is broadcast to other GPUs, constituting an \texttt{AllGather} operation with the same time complexity as \texttt{ReduceScatter}. The overall time complexity of the ring-based \texttt{AllReduce} is given by $t_{ar}=2(p-1)(\alpha+\frac{v}{w \times p})$.

However, predicting collective communication time with high accuracy using the $\alpha-\beta$ cost model is challenging in certain scenarios. First, in addition to the ring algorithm, NCCL introduces new communication algorithms like Tree \cite{TreeComm}, Collnet, CollnetDirect, and CollnetChain. Consequently, a single cost model struggles to formulate the communication time for all these algorithms. Second, In-Network Aggregation solutions are widely implemented in production GPU clusters. These solutions offload \texttt{AllReduce} onto network switches to accelerate and scale distributed training \cite{graham2020scalable} \cite{li2023a2tp} \cite{liu2023network}. 

In this work, we adopt a straightforward yet effective profiling-based approach to model the performance of collective communication. Specifically, we utilize 
$$t(o,v,p^0 \times p^1) = v/w(o,v,p^0 \times p^1)$$
to evaluate the time consumption of a collective communication operator ($o$) with a given data size ($v$) and a specified participant GPU device mesh ($p^0 \times p^1$, where $p^0$ denotes the number of GPUs in a node, and $p^1$ is the number of nodes). Here, $w(o,v,p^0 \times p^1)$ represents the effective bandwidth obtained through performance profiling on the target GPU cluster in advance, as illustrated in Figure \ref{fig: commu_bench}. In cases where $v$ is not profiled, we employ the interpolation method to predict the effective bandwidth and communication time. This work focuses on four key collective communication operations: \texttt{AllReduce(AR)}, \texttt{AllGather(AG)},  \texttt{ReduceScatter(RS)} and \texttt{Broadcast(BC)}.

\begin{figure*}[t]
    \centering
    \includegraphics[width=\linewidth]{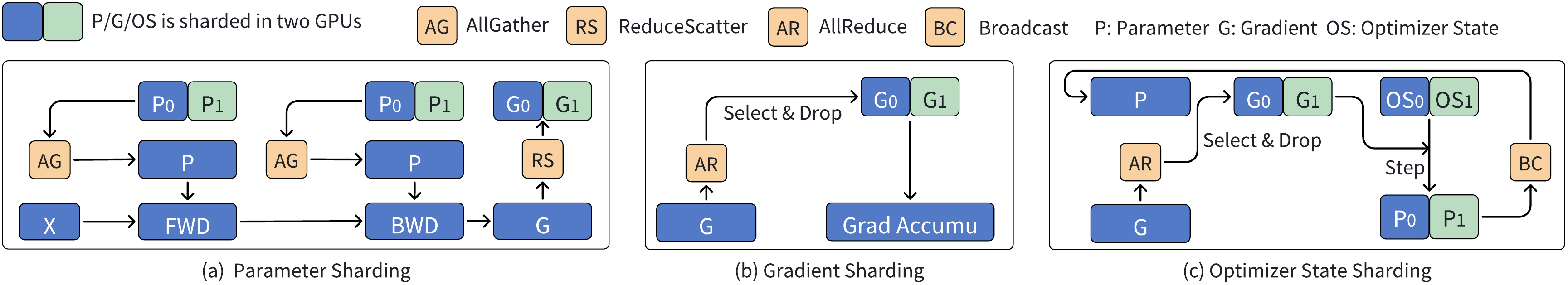}
    \caption{Analysis of inserted collective communication operations when individually sharding parameters, gradients, and optimizer states.}
    \label{Fig: lins_partition}
\end{figure*}

\subsection{Flexible Model States Sharding with Dependency Rule}
\label{subsec:flexibleModelStatePar}

We adopt three sharding strategies, namely \emph{Full-Replica}, \emph{Full-Sharding}, and \emph{Partial-Sharding}, and provide the flexibility for each of the three components within the model states (\emph{P}, \emph{G}, and \emph{OS}) to independently select a sharding strategy. To encapsulate these strategies, we introduce sharding factors $s_p = s_p^0 \times s_p^1$, $s_g = s_g^0 \times s_g^1$ and $s_{os} = s_{os}^0 \times s_{os}^1$, representing the number of GPUs and the device mesh over which the tensors of \emph{P}, \emph{G}, and \emph{OS} are sharded, respectively.
Setting the factor to 1 implies full replication of the tensor, simplifying to vanilla data parallelism if all components (\emph{P}, \emph{G}, and \emph{OS}) choose the \emph{Full-Replica} strategy. Conversely, setting the factor equal to the DP size results in complete tensor sharding, with each GPU holding ${1}/{s_{dp}}$ of the tensor. For instance, in ZeRO-3, all components ($P$, $G$, and $OS$) choose the \emph{Full-Sharding} strategy.
\emph{Partial-Sharding} emerges when the factor falls between 1 and $s_{dp}$, indicating tensor sharding across a subgroup of GPUs and replication across subgroups.

A dependency rule is crucial when flexibly sharding the model states to avoid unnecessary data movement and storage. Throughout the training, the framework employs local parameters for gradient computation and synchronized gradients for updating local optimizer states. If a GPU oversees extra gradients or optimizer states unrelated to its local parameters, launching additional communication becomes necessary. This incurs significant and avoidable expenses. Figure \ref{fig: dependencebased} illustrates an instance of this scenario, highlighting the impact when setting $s_p=s_g=2, s_{os}=1$. Before independently sharding $P$, $G$, and $OS$, we establish the following constraints:
$$ R \geq s_{dp}^0 \geq s_{os}^0 \geq s_{g}^0 \geq s_{p}^0,  \quad N \geq s_{dp}^1 \geq s_{os}^1 \geq s_{g}^1 \geq s_{p}^1,$$
where $s_{dp}^0$ and $s_{dp}^1$ is the device mesh of DP ranks, $R$ denotes the GPU count per node, and $N$ is the node number. As shown in Figure \ref{fig: dependencebased}, adhering to the dependency avoids unnecessary data movement and storage.

\subsection{Communication Time Analysis}

In this subsection, we analyze the communication cost associated with individually partitioning parameters, gradients, and optimizer states. Figure \ref{Fig: lins_partition} provides an overview of the inserted collective communication operations for each component.

\subsubsection{Parameters Sharding}
\label{subsec:parameterPartitioning}
When $s_p^0 \times s_p^1 = s_p > 1$, the $\Phi$ parameters of a model are split into $s_p$ shards, with each GPU managing one shard. As shown in Figure \ref{Fig: lins_partition}(a), during each forward and backward pass of every micro-batch in a step, the training system orchestrates the collection of parameters shards from other GPUs to reconstruct the complete set of model weights required for computations. This is achieved using \texttt{AllGather} on $s_p$ GPUs.
In each micro-batch of a step, after the gradients are generated during the backward phase, the training framework launches \texttt{ReduceScatter} to aggregate and distribute gradients across $s_p$ GPUs. 
The training framework performs \texttt{AllGather} and \texttt{ReduceScatter} at the granularity of a module within a Transformer layer. The input size for $i$-th module of a layer is $2\Phi_i$  (using FP16). 
The communication time attributable to parameters sharding for $M$ micro-batches of a step is given by:
\[ T_p = ML \sum_{i=0}^{K} \left(2 t(\texttt{AG}, 2\Phi_i, s_p^0 \times s_p^1) + t(\texttt{RS}, 2\Phi_i, s_p^0 \times s_p^1)\right), \]
where $L$ denotes the number of layers and $K$ is the number of modules of a layer.

\begin{figure}[t]
    \centering
    \includegraphics[width=1\linewidth]{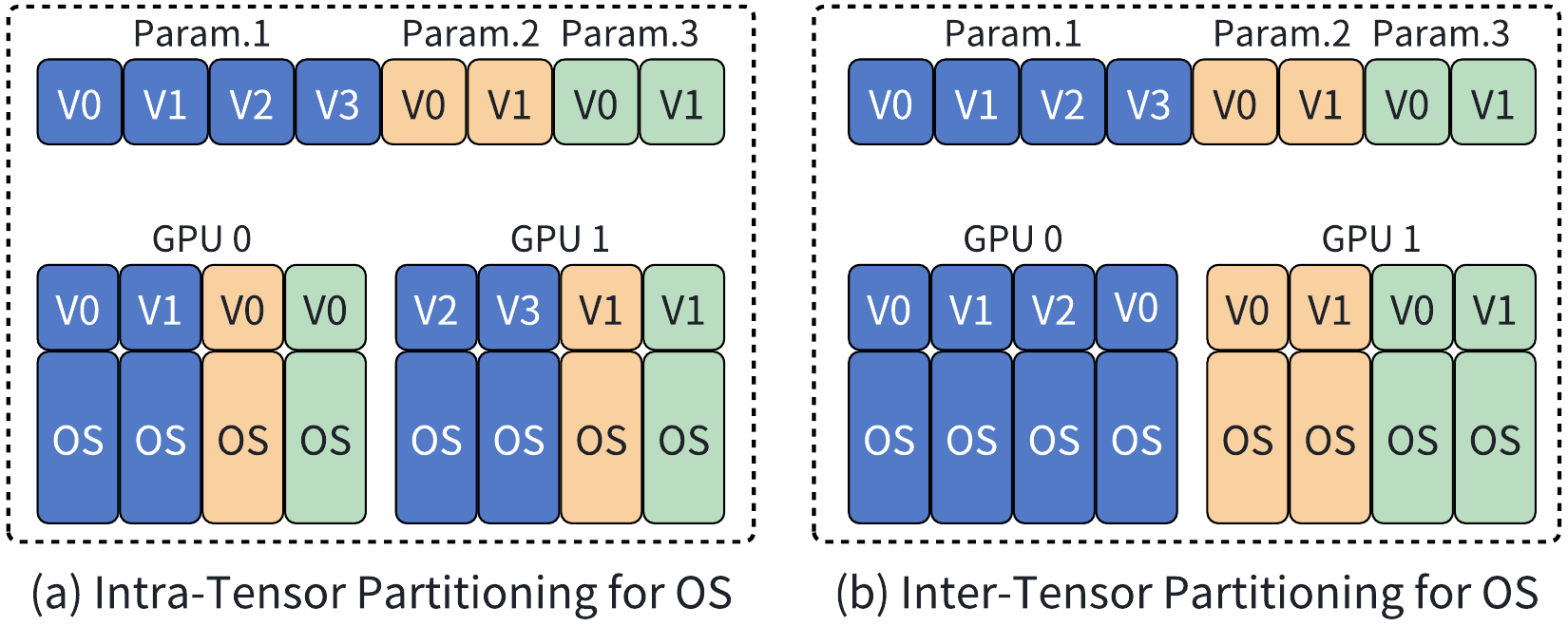}
    \caption{Sharding scheme for optimizer states. (a) shards each optimizer states tensor into multiple devices along with the corresponding optimizer states. (b) distributes each tensor of optimizer states in its entirety.}
    \label{fig:partitionScheme}
\end{figure}

Taking Figure \ref{fig: dependencebased} (a) as an example, when $s_p^0 \times s_p^1=2 \times 1$,  \texttt{AllGather} and \texttt{ReduceScatter}  are executed within the same node. parameters sharding allows for overlapped communication with computation. During the forward or backward computation of a module, it is feasible to execute \texttt{AllGather} and \texttt{ReduceScatter} for the subsequent module.

\subsubsection{Optimizer States Sharding}

When $s_{os}^0 \times s_{os}^1 = s_{os} > 1$, a total of $s_{os}$ GPUs collectively possess a complete duplicate of optimizer states. Following parameters sharding with $s_{p}$, each parameters is stored and replicated across $s_{dp}/s_{p}$ GPUs. In this configuration, optimizer states may exhibit redundancy, with $s_{dp}/s_{p}$ replicas distributed across the cluster.
To reduce this redundancy, we introduce a solution by allowing $s_{os} > s_{p}$, affording flexibility to reduce redundancy. In this scenario, the optimizer states for $\Phi/s_{p}$ parameters are shared by $s_{os}/s_{p}$ GPUs, forming an optimizer states sharding subgroup. Illustrated in Figure \ref{fig: dependencebased} (b), GPU-0 and GPU-2 share common parameters shards but maintain distinct optimizer states shards, forming an optimizer states sharding subgroup.

After the backward pass of the last micro-batch in a step, each GPU updates $\Phi/s_{os}$ parameters based on the optimizer states. Before the update phase, each GPU should gather and aggregate gradients for parameters within its optimizer states. 
To optimize this process, we employ  \texttt{AllReduce}  on gradients across $s_{dp}/s_{p}$ GPUs sharing the same set of parameters (in the amount of $\Phi/s_{p}$), as illustrated in Figure \ref{Fig: lins_partition} (c). In Figure \ref{fig: dependencebased} (b), we execute \texttt{AllReduce} on GPU-0/2/4/6. Given that $s_{os} > s_{p}$, each GPU receives additional gradients not managed by its optimizer states. To resolve this issue, we employ a \emph{select \& drop} mechanism, enabling each GPU to exclusively select the necessary gradients from the output of the \texttt{AllReduce}.
% \todo{say some words about select and drop and why need it}

Following the completion of parameters updates, it is essential to spread updated parameters among GPUs within the same optimizer states sharding subgroup. For example, In Figure \ref{fig: dependencebased} (b), GPU-0 should send its updated parameters to GPU-2. The choice of the collective communication primitive relies on the sharding scheme employed for optimizer states. Typically, two mechanisms govern the sharding of optimizer states across GPUs, as illustrated in Figure \ref{fig:partitionScheme}. 1) The \emph{intra-tensor} approach involves evenly splitting a single parameters along with its states into multiple shards, which are then distributed to different GPUs. In this scenario,  \texttt{AllGather}  proves effective, ensuring that each GPU receives all updated values for parameters stored in its local memory. 2) The \emph{inter-tensor} approach distributes each parameters and its states as a whole across devices, using a greedy algorithm to balance GPU memory usage. In this case, direct usage of \texttt{AllGather} is not feasible, as each GPU may not have an identical number of updated parameters. To address this, the updated parameters can be spread by broadcasting each shard separately using an NCCL group call. 
% An alternative  approach involves padding the partition to the same size and performing \texttt{AllGather}.

While both sharding schemes remain compatible with mix-precision training using FP16, the inter-tensor approach is recommended for FP8 training \cite{FP8Training}. This preference is pivotal since the distribution of per-tensor scaling factors becomes imperative when dealing with FP8 shards. Consequently, we adopt the inter-tensor approach in this work. Following the optimizer update stage, a series of \texttt{Broadcast} operations are initiated on $s_{os}/s_{p}$ GPUs to disseminate updated parameters.

The training system launches $2\Phi/U$ \texttt{AllReduce} operations with a specified bucket size $U$ to synchronize gradients for a model with $\Phi$ trainable parameters (utilizing FP16). The \texttt{AllReduce} communication time attributed to optimizer states sharding for a given step is expressed as:
\[ T_{os}^{0} = \frac{2\Phi}{Us_{p}} \left( t(\texttt{AR}, U, \frac{s_{dp}^0}{s_p^0} \times \frac{s_{dp}^1}{s_p^1}) \right). \]
For disseminating updated parameters, the training system utilizes a group of \texttt{Broadcast} operations. Each \texttt{Broadcast} operation processes an input of size $2\Phi/s_{os}$ on average, and the system executes $s_{os}/s_p$ such operations. The \texttt{Broadcast} communication time attributable to optimizer states sharding during a step is given by:
\[ T_{os}^1 = \frac{s_{os}}{s_{p}} \left( t(\texttt{BC}, \frac{2\Phi}{s_{os}}, \frac{s_{os}^0}{s_{p}^0} \times \frac{s_{os}^1}{s_{p}^1}) \right). \]
optimizer states sharding facilitates the potential for overlapped communication and computation. During the backward computation of the $i$-th layer, we can perform \texttt{AllReduce} on gradients generated on layer $i+1$. Simultaneously, during the forward computation of the $i$-th layer, it is also possible to broadcast the latest parameters (updated in the previous step) for the next layer.

\subsubsection{Gradients Sharding}
When $s_{g}^0 \times s_{g}^1 = s_{g} > 1$, a total of $s_{g}$ GPUs collectively hold a complete copy of gradients generated at each micro-batch of every step. As depicted in Figure \ref{fig: dependencebased}, if $s_{g}=s_{p}$, each GPU retains $\Phi/s_{p}$ gradients, accumulating them at every micro-batch based on the parameter sharding mechanism. In this work, we introduce the flexibility to set $s_{g} > s_{p}$ to conserve GPU memory. For simplicity, we impose the following constraints on the selection of $s_g$:
\[ s_g \in \{s_p, s_{os}\}. \]

In the scenario where $s_g > s_p$, each GPU initiates an \texttt{AllReduce} operation on $s_{g}/s_{p}$ GPUs to aggregate and distribute gradients in every micro-batch, excluding the last one. Following the aggregation, each GPU retains only the gradients allocated to it, discarding the surplus. For instance, in Figure \ref{fig: dependencebased} (c), GPU-0 and GPU-2 can employ such a \emph{select \& drop} mechanism to shard gradients.
It is noteworthy that alternative approaches might leverage \texttt{ReduceScatter} to achieve similar outcomes. However, the \emph{inter-tensor} approach in sharding optimizer states leads to uneven shard sizes per GPU, making \texttt{ReduceScatter} less suitable. Thus, we opt for \texttt{AllReduce} on gradients, preserving only the relevant ones.
Assuming  \texttt{AllReduce} is executed with bucket size $U$, the communication time attributable to gradients sharding of a step can be expressed as: 
\[ T_g = (M-1)\frac{2\Phi}{Us_{p}}  \left( t(\texttt{AR}, U, \frac{s_{g}^0}{s_p^0} \times \frac{s_{g}^1}{s_p^1}) \right). \]
Gradients sharding can also overlap communication with computation. During the backward computation for  $i$-th layer, we can concurrently execute \texttt{AllReduce} for  $(i+1)$-th layer.

\subsubsection{Summary}

Based on the aforementioned analysis, we can conclude that the single-step communication time of distributed LLM training with a flexible model states sharding strategy is the sum of three components:
\[ T_{comm}(s_p^0, s_p^1, s_g^0, s_g^1, s_{os}^0, s_{os}^1) = T_p + T_g + T_{os}^0 + T_{os}^1. \]

\subsection{GPU Memory Consumption Analysis}

In the context of mixed-precision training with the Adam optimizer, the GPU memory consumed by model states during training can be expressed as the sum of memory allocated for sharded parameters, gradients, and optimization states:
\[ D_{modelstate}(s_p^0, s_p^1, s_g^0, s_g^1, s_{os}^0, s_{os}^1) = \frac{2\Phi}{s_p^0 s_p^1} + \frac{2\Phi}{s_g^0 s_g^1} + \frac{12\Phi}{s_{os}^0 s_{os}^1}. \]
Additionally, the total GPU memory consumption,  $D_{total}$, can be encapsulated by:
\[ D_{total} = D_{modelstate} + D_{activation} + D_{tmp}, \]
where \(D_{{activation}}\) is the memory consumed by activations during training, and \(D_{{tmp}}\) denotes the temporary memory used by communication buffers or other transient variables. Existing methodologies \cite{Megatron-LM} \cite{ZeRO} for analyzing and predicting activation memory usage are seamlessly integrated into our present study.

\section{System Design \& Communication Overlap}
\label{sec_design}

To reduce the communication overhead of ZeRO for efficient  LLM training, we introduce \SysName. It leverages an expanded model states sharding space and is adept at identifying the most communication-efficient  factors. We focus on how \SysName systematically optimizes the training performance with a flexible model states sharding strategy.

\subsection{System Architecture}

\begin{figure}[t]
    \centering
    \includegraphics[width=\linewidth]{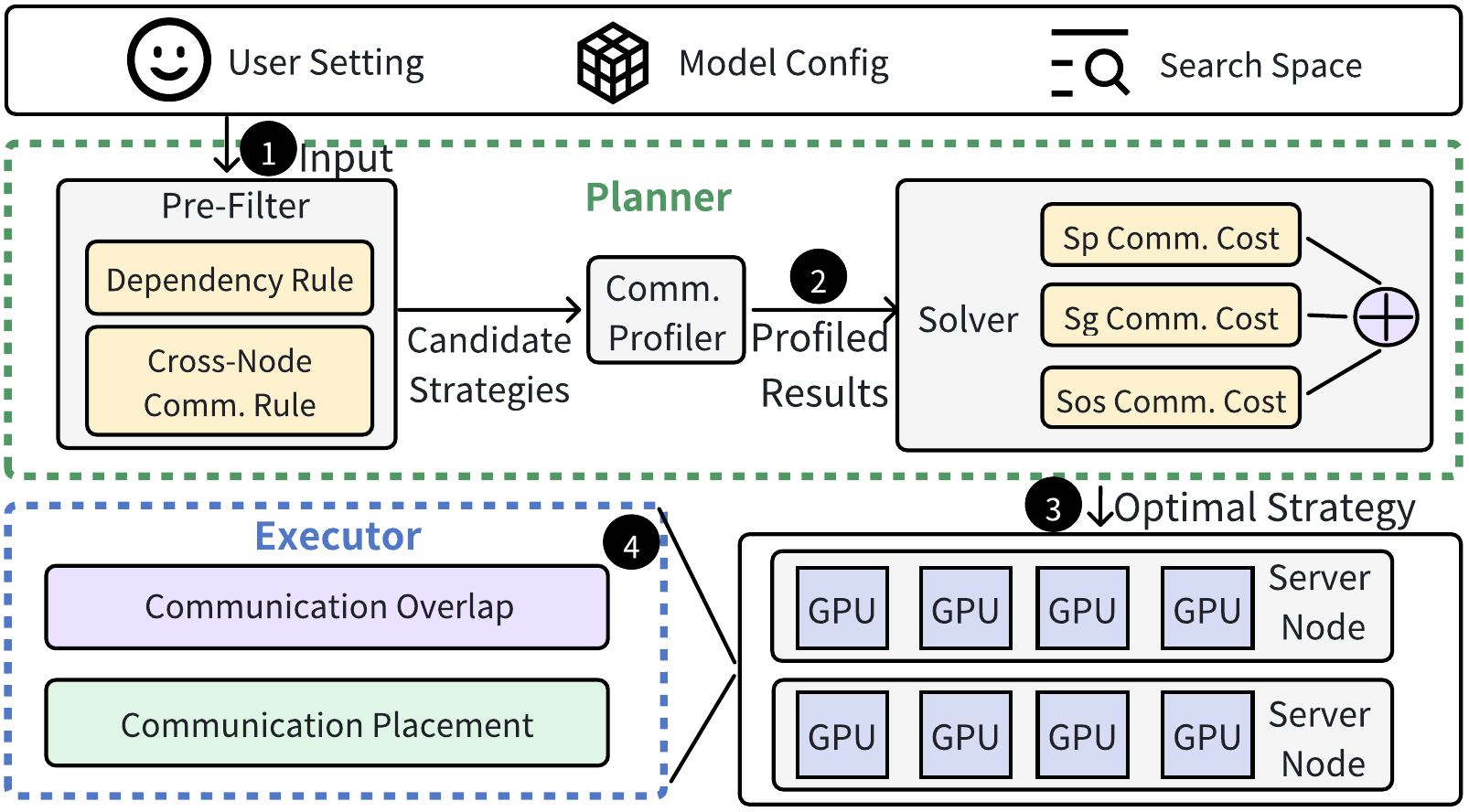}
    \caption{Overview of \SysName architecture and workflow. The Planner identifies the optimal solution for model states sharding. The Executor executes LLM training using the selected strategy, and enhances communication performance through overlap and placement optimization. }
    \label{fig:framework}
\end{figure}

Figure \ref{fig:framework} illustrates the two components of \SysName: the Planner and the Executor. (1) The Planner identifies the optimal solution for model states sharding. This component integrates three modules: the Pre-Filter, narrowing the search space based on specific rules; the Communication-Profiler, offering predictions for collective communication time; and the Solver, constructed by a memory and communication cost model to identify the optimal strategy. (2) The Executor is accountable for executing LLM training using the selected strategy. This component integrates two essential modules to enhance communication efficiency. The Communication Overlap strategy offers a fine-grained overlap for computation and communication. Moreover, \SysName employs the topology-aware Communication Placement strategy to reduce network communication across spine switches, enhancing overall efficiency.

\textbf{Workflow.} 
\circled{1} \SysName begins by having users define LLM architecture, specifying metadata such as layer number and sequence length, along with hyper-parameters like micro-batch size and micro-batch number. Users also provide settings for the training cluster, including the total number of GPUs and GPU memory capacity. \circled{2} The Planner eliminates certain strategies that may incur additional communication costs, resulting in a set of alternative strategies. The Communication-Profiler, operating offline, provides communication time data, aiding the Planner in estimating step time for these alternatives. \circled{3} Using an optimization problem solver, the Planner identifies the optimal strategy. \circled{4} Subsequently, the Executor runs the training job using the chosen strategy, enhanced by Communication Overlap and topology-aware Communication Placement strategy.

\begin{figure}[t]
    \centering
    \includegraphics[width=\linewidth]{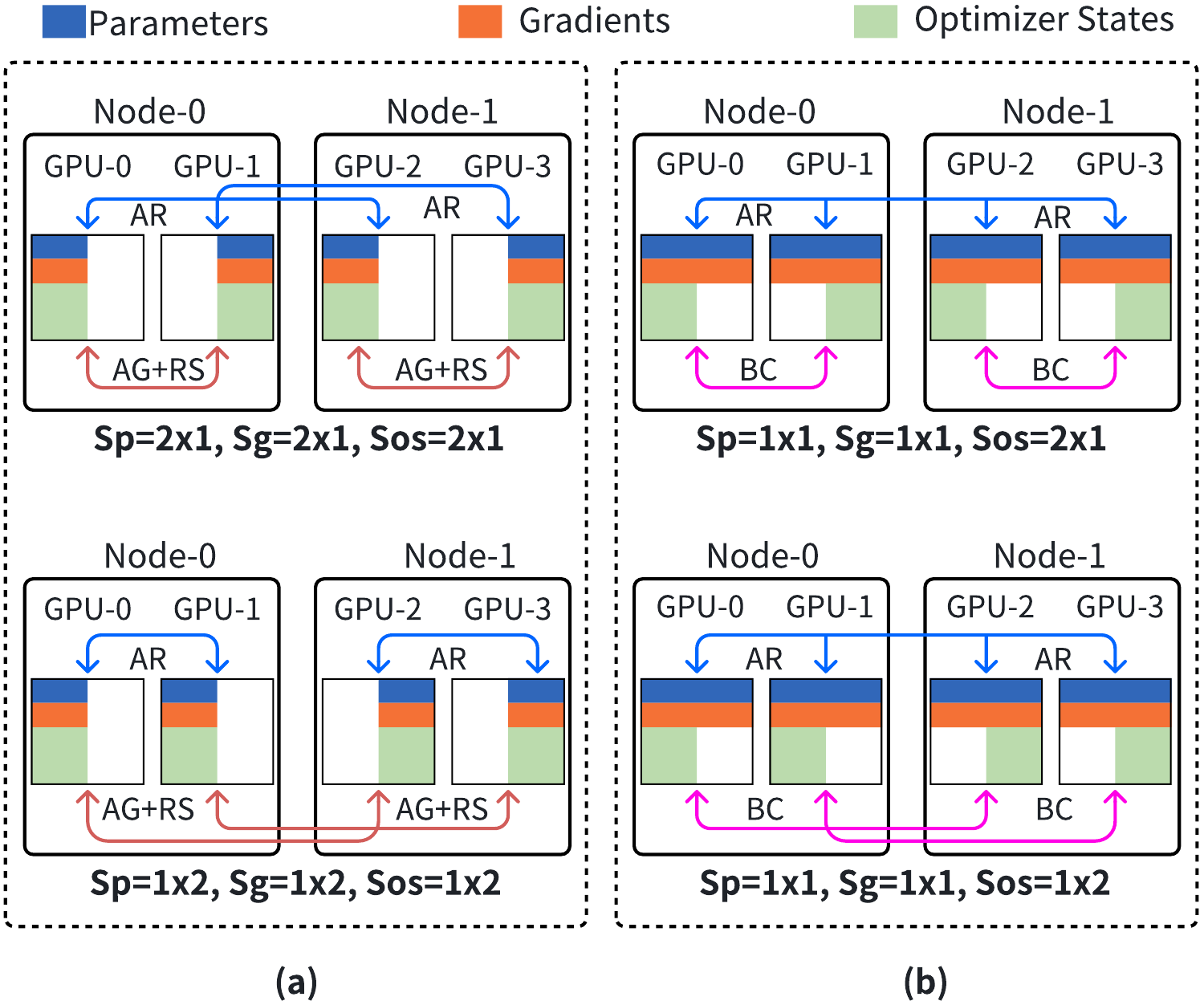}
    \caption{Example of sharding model states within the same node. In (a) assigning $s_p^0=1, s_p^1=2$ results in increased cross-node communication caused by  \texttt{AllGather} and \texttt{ReduceScatter}. In (b) with $s_{os}=2$, setting $s_{os}^0=1, s_{os}^1=2$ creates cross-node \texttt{Broadcast}.}
    \label{fig:takeaway}
\end{figure}

\SysName utilizes real-system profiling to ascertain the effective bandwidth of three used collective communication operations (i.e., \texttt{AllGather}, \texttt{ReduceScatter}, \texttt{AllReduce} and  \texttt{Broadcast}) across diverse communication sizes and device meshes. Consequently, \SysName estimates the communication latency induced by model states sharding.

\subsection{Execution Planner}

The Execution Planner generates the optimal combination strategy for the input model with the provided hardware information. \SysName formulates an optimization problem to search for the optimal $\{ s_p^0, s_p^1, s_g^0, s_g^1, s_{os}^0, s_{os}^1 \}$ by minimizing the sum of communication costs subject to memory constraints. The integer programming problem is defined as follows:
\begin{align}
     \text{Minimize} \quad & T_{comm}(s_p^0, s_p^1, s_g^0, s_g^1, s_{os}^0, s_{os}^1)  \label{eq:objective} \\
     \text{Subject to} \quad &  D_{total} \leq \texttt{GPU\_Memory\_Capacity} \label{eq:capacity} \\
    & 1 \leq s_p^0 \leq s_g^0 \leq s_{os}^0 \leq s_{dp}^0 \leq R  \label{eq:rule1}  \\
    & 1 \leq s_p^1 \leq s_g^1 \leq s_{os}^1 \leq s_{dp}^1 \leq N \label{eq:rule2} \\
    & s_{i}^0 \times k = s_{dp}^0, \,\, k \in \mathbb{Z}, \quad  i \in \{p, g, os\} \label{eq:constrain1} \\
    & s_{j}^1 \times k =  s_{dp}^1, \,\,  k \in \mathbb{Z},  \quad  j \in \{p, g, os\} \label{eq:constrain2} \\
    & s_{i}^0 = s_{dp}^0, \quad \text{if } s_{i}^1 > 1, \quad  i \in \{p, g, os\} \label{eq:rule3} 
    % \\
    % & s_{j}^0 \leq s_{dp}^1, \quad \text{if } s_{j}^1 = 1, \quad  j \in \{p, g, os\} \label{eq:rule4}
\end{align}

This problem minimizes the communication cost of LLM training with respect to the GPU memory capacity (Equation \ref{eq:capacity}) and the dependency rules outlined in Section \ref{subsec:flexibleModelStatePar} (Equation \ref{eq:rule1}, \ref{eq:rule2}).  
Instead of exhaustively iterating through all possible solutions for $ \{s_p^0, s_p^1, s_g^0, s_g^1, s_{os}^0, s_{os}^1 \}$, we optimize the efficiency of assignment strategy exploration by incorporating two filters. Firstly, $s_{dp}^{0}$ should be divisible by $s^0_{{p,g,os}}$ (Equation \ref{eq:constrain1}), ensuring the participation of all GPUs within a node in the training process. Additionally, $s_{dp}^{1}$ should be divisible by $s^1_{{p,g,os}}$ (Equation \ref{eq:constrain2}), allowing for the utilization of all nodes in the training.
Secondly, to minimize cross-node communication (Equation \ref{eq:rule3}), a priority is placed on employing fewer nodes when sharding $P$, $G$, and $OS$ independently. For instance, in Figure \ref{fig:takeaway}(a), selecting $(s_p^0=1, s_p^1=2)$ necessitates launching \texttt{AllGather} and \texttt{ReduceScatter} on two nodes every micro-batch, while opting for $(s_p^0=2, s_p^1=1)$ results in reduced cross-node communication costs. In Figure \ref{fig:takeaway}(b), setting $(s_{os}^0=1, s_{os}^1=2)$ induces cross-node \texttt{AllGather} for spreading updated parameters per step, whereas $(s_{os}^0=2, s_{os}^1=1)$ confines this communication within a node.
Based on these filters, \SysName can efficiently employ a brute-force search method to obtain the optimal solution, effectively navigating the solution space with reduced complexity.

\subsection{Computation-Communication Overlap}

\SysName uses specific hooks of PyTorch 2.1, as detailed in Table \ref{tab:hooks}, to facilitate the necessary NCCL  communications for sharding $P$, $G$, and $OS$. Timely initiation of these operations is important for ensuring both correctness and efficiency. 
% In our implementation, we opted for a group of \texttt{Broadcast} operations for optimizer updates on each GPU, favoring simplicity with consistent time complexity. 

\renewcommand{\arraystretch}{0.95}
\begin{table}[h]
    \centering
    \caption{Hooks used by \texttt{\SysName}}
    \label{tab:hooks}
    \begin{tabular}{ll}
        \toprule
        Module & Hook Name \\
        \midrule
        \texttt{torch.nn.modules} & \texttt{register\_forward\_hook} \\
        \texttt{torch.nn.modules} & \texttt{register\_forward\_pre\_hook} \\
        \texttt{torch.nn.modules} & \texttt{register\_full\_backward\_hook} \\
        \texttt{torch.nn.modules} & \texttt{register\_full\_backward\_pre\_hook} \\
        \texttt{torch.Tensor} & \texttt{register\_hook} \\
        \bottomrule
    \end{tabular}
\end{table}
\renewcommand{\arraystretch}{1}

\subsubsection{Overlap for Parameters Sharding}

\begin{figure}[t]
    \centering
    \includegraphics[width=1\linewidth]{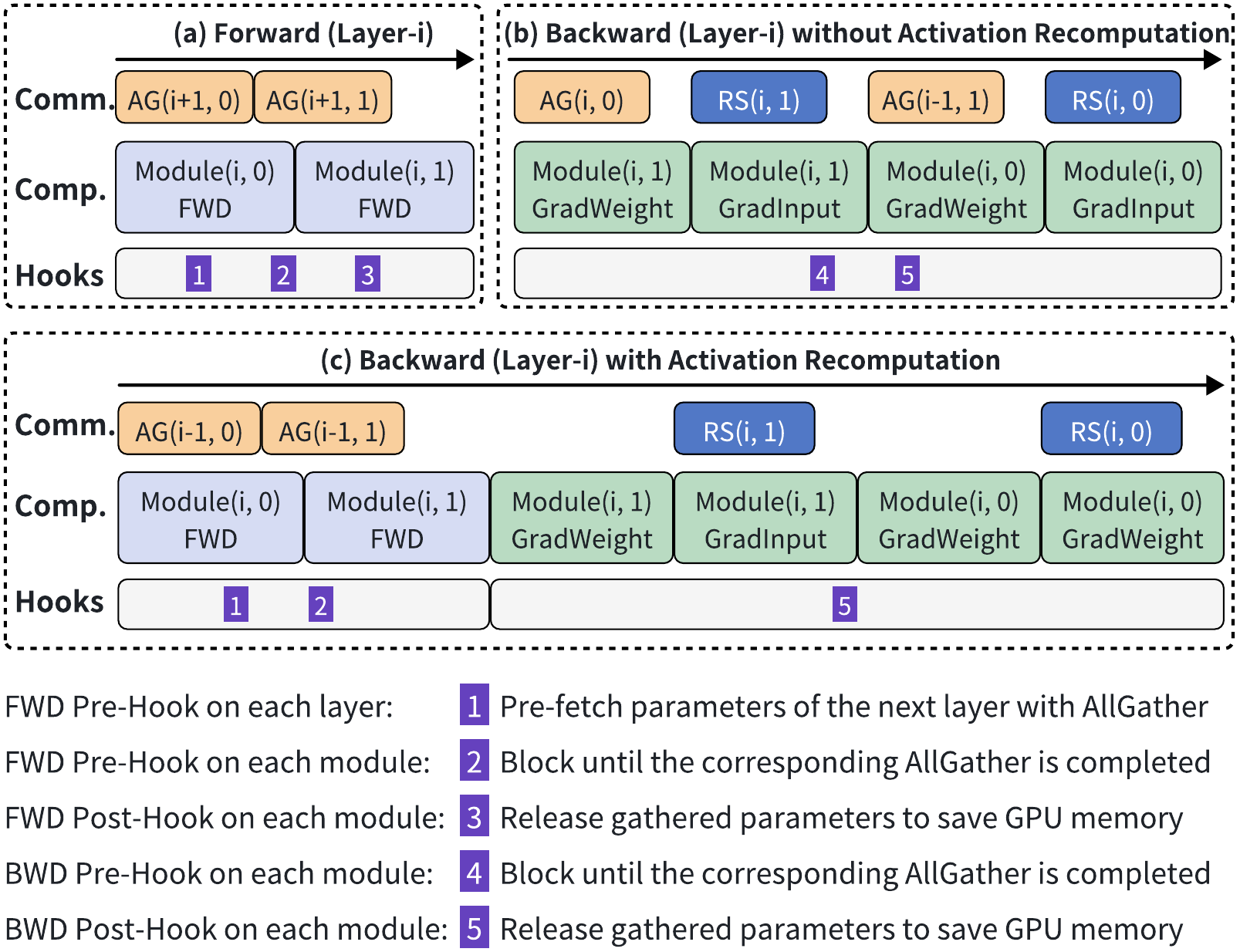}
    \caption{Overlapping strategy for parameters sharding and corresponding control hooks. (a) illustrates the concurrent execution of \texttt{AllGather} for pre-fetching parameters  with the forward computation. (b) presents the overlap of \texttt{AllGather} and \texttt{ReduceScatter} with backward computation. (c) demonstrates the communiction-computation overlap strategy during the backward phase when activation recomputation is enabled.}
    \label{fig:overlap}
\end{figure}

As shown in Figure \ref{fig:overlap} (a), prior to initiating the forward computation in a  layer, a sequence of \texttt{AllGather} operations is launched to proactively fetch parameters for each module (such as linear modules) of the subsequent layer. This strategic prefetching enables \SysName to seamlessly overlap the computation of the current layer with communication tasks for the next layer, enhancing overall efficiency. To coordinate this process for each module, \SysName leverages  \texttt{register\_forward\_pre\_hook} to await the completion of the corresponding \texttt{AllGather}. Upon concluding the forward computation for a module, \SysName releases gathered parameters, triggered by \texttt{register\_forward\_hook}.

In the backward pass, each module within a layer computes gradients for both its weights (\texttt{GradWeight}) and its input (\texttt{GradInput}). After the computation of \texttt{GradWeight}, \SysName initiates \texttt{ReduceScatter} on it for gradients distribution and synchronization. In scenarios without activation recomputation, launching all \texttt{AllGather} operations simultaneously to fetch parameters of modules in the next layer is not optimal. This is because such operations would obstruct the execution of \texttt{ReduceScatter}, leading to sub-optimal training performance.
To address this challenge, we adopt a more strategic approach by fetching parameters of only the next module, triggered by \texttt{register\_backward\_pre\_hook}. Consequently, as shown in Figure \ref{fig:overlap} (b), \SysName efficiently overlaps  \texttt{AllGather} with the computation of \texttt{GradWeight}.  \SysName also overlaps \texttt{ReduceScatter} with the computation of \texttt{GradInput}. 
Furthermore, we decouple the life-cycle of \texttt{ReduceScatter} from the backward function. In cases where the \texttt{ReduceScatter} operation is not completed by the time \texttt{GradInput} computation concludes, \SysName seamlessly proceeds to the backward computation of the next module instead of causing unnecessary blocking. 

As we decouple the life-cycle of \texttt{ReduceScatter} from the backward function, the completion of the backward function does not guarantee the availability of the true gradients for the corresponding weights. Therefore, an additional post-hook is applied to the AccumulateGrad of each parameter, ensuring the completion of the corresponding \texttt{ReduceScatter} operation before utilizing these gradients for subsequent communication or computation tasks. Examples of such tasks include launching \texttt{AllReduce} operations on this data, facilitating gradient synchronization across data parallelism ranks.

In Figure \ref{fig:overlap} (c), with the activation recomputation mechanism enabled, \SysName requires an additional forward computation for a layer during the backward pass. At the beginning of this secondary forward pass for a layer, \SysName initiates a series of \texttt{AllGather} operations to proactively fetch parameters for the subsequent layer. Importantly, in contrast to the first forward phase, \SysName retains the gathered parameters after the secondary forward computation for subsequent gradients computation.
Following the computation of \texttt{GradWeight}, \SysName proceeds to initiate a \texttt{ReduceScatter} operation on it. \SysName  efficiently overlaps both \texttt{AllGather} and \texttt{ReduceScatter} operations with computation during the backward phase.

\subsubsection{Overlap for Optimizer States Sharding}

During the backward phase of the last micro-batch, each GPU aggregates gradients for parameters within its optimizer states by using \texttt{AllReduce} across data parallelism ranks. To optimize this process, we implement a bucketing strategy  where a parameter is placed in a bucket after its backward computation \cite{PyTorchDistributed}. Once a bucket is full, all gradients of the parameters within it are flattened into a contiguous buffer. Then, we execute \texttt{AllReduce} on this buffer without blocking the backward computation of the remaining layers. This approach is triggered by hooks on the AccumulateGrad of parameters.

We utilize the asynchronous communication mechanism of NCCL to overlap parameters broadcast with the forward computation of the next step. It is crucial to ensure that all parameters have updated values before being used for both computation and communication. We implement two optimizations to address this challenge. Firstly, we register a hook for each module by using  \texttt{register\_forward\_pre\_hook}, ensuring that the parameters to be used are the most recent ones before any computation or communication takes place. Secondly, we  manage the order of forward computation to align with the sequence of parameters broadcast. This synchronization ensures that the parameters used in the forward computation are the ones that have been successfully broadcasted, avoiding unnecessary blocking during the forward phase.

\subsubsection{Overlap for Gradients Sharding}
In the case of $s_g > s_p$, the system initiates \texttt{AllReduce} operations on $s_g/s_p$ GPUs for gradients aggregation and distribution in each micro-batch excluding the final one. To further enhance efficiency, a bucketing strategy, similar to the one employed for optimizer states sharding, is leveraged. The  \texttt{AllReduce} communication process seamlessly overlaps with the backward computation, facilitated through hooks integrated into the AccumulateGrad of parameters. Following the \texttt{AllReduce}, a GPU rank only retains gradients pertinent to it, releasing other gradients to conserve GPU memory.

\begin{figure}[t]
    \centering
    \includegraphics[width=\linewidth]{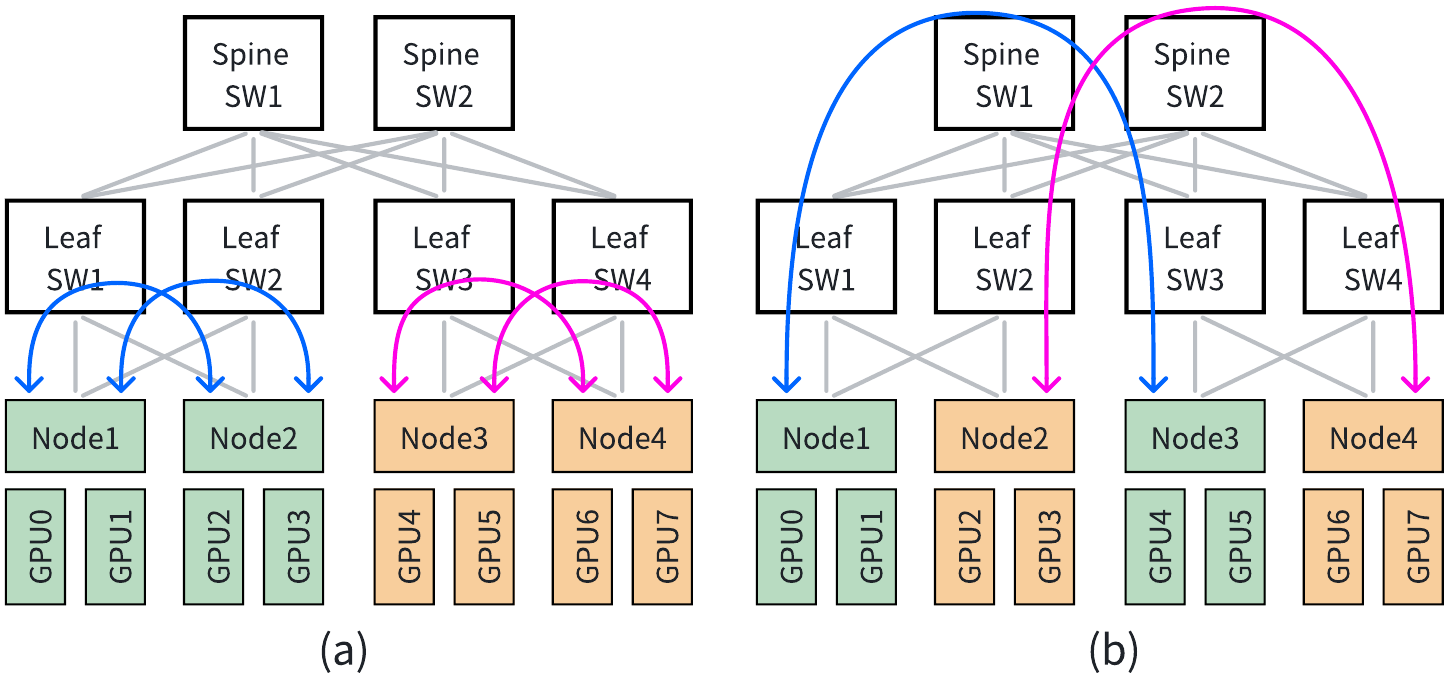}
    \caption{Examples of communication placement. When $s^1_{p,g,os} > 1$, (a) depicts an instance where groups of $s^1_{p,g,os}$ nodes are organized under the same spine switch, facilitating intra-switch communication, which is the preferred approach by \SysName. (b) showcases a scenario in which the groups of $s^1_{p,g,os}$ nodes are distributed across different spine switches, necessitating extensive cross-switch communication.}
    \label{fig:placement}
\end{figure}

\subsection{Communication Placement}

To further improve training performance, \SysName tries to minimize communication traffic across spine-switches within the leaf-spine network architecture, which is commonly used in current GPU data centers. Typically, GPU nodes are connected to leaf-switches, and these leaf-switches are interconnected by spine-switches. When GPUs are not under the same leaf switch, communication has to go through spine switches, leading to increased latency and potential network congestion. In this study, when  $s^1_{p,g,os}>1$, \SysName aims to optimize collective communications by utilizing fewer leaf switches as illustrated in Figure \ref{fig:placement} (a), instead of incurring extensive cross-switch communication as shown in Figure \ref{fig:placement} (b). For instance, with  $(s_{p}^0=8, s_{p}^1=4)$, \SysName strategically groups four nodes under the same leaf switches to perform collective communications like \texttt{AllGather} and \texttt{ReduceScatter}. This strategic approach aims to minimize inter-switch communication latency, thereby reducing the additional communication overhead introduced by the sharding of model states.

%-------------------------------------------------------------------------------

\begin{figure}[t]
    \centering
    \includegraphics[width=\linewidth]{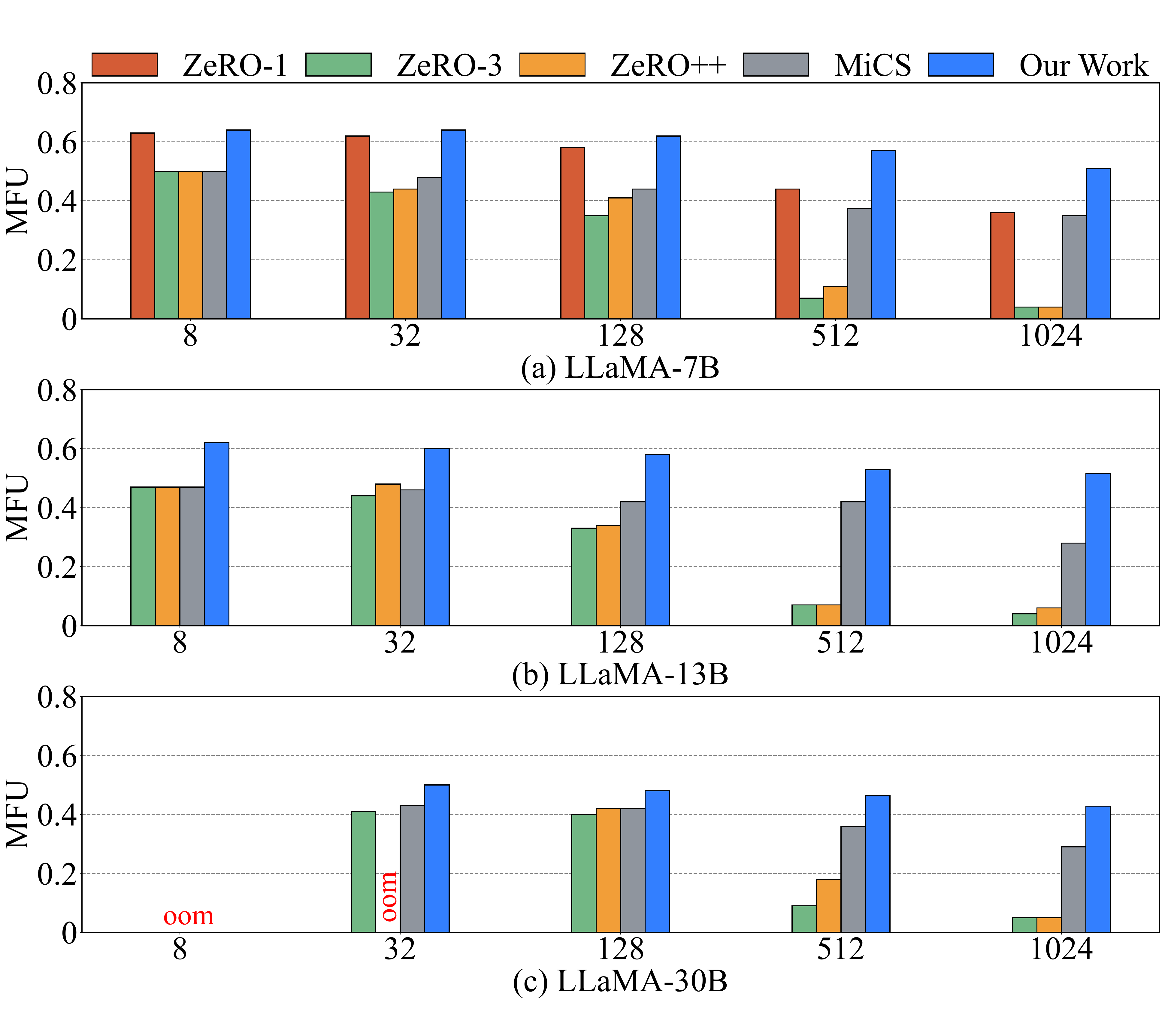}
    \caption{End-to-end evaluation results (MFU) of training LLaMA-based Models from 8 GPUs to 1024 GPUs.}
    \label{e2eMFU}
\end{figure}

\renewcommand{\arraystretch}{0.95}
\begin{table}%[htbp]
    \centering
    \caption{Strategies used for sharding $P$, $G$ and $OS$.}
    \label{tab:configs}
    \begin{threeparttable}
        \begin{tabular}{l|l|cccccc}
            \toprule
            {Model} & {Approach\tnote{1}} & {$s_p^0$} & {$s_p^1$} & {$s_g^0$} & {$s_g^1$} & {$s_{os}^0$} & {$s_{os}^1$} \\ \midrule
            \multirow{6}{*}{LLaMA-7B} & Our Work & 1 & 1 & 1 & 1 & 8 & 1 \\
            & ZeRO-1 & 1 & 1 & 1 & 1 & R & N \\
            & ZeRO-3 & R & N & R & N & R & N \\
            & MiCS & 8 & 1 & 8 & 1 & 8 & 1 \\
            & ZeRO++\tnote{2} & R & N & R & N & R & N \\ \midrule
            \multirow{5}{*}{LLaMA-13B} & Our Work & 4 & 1 & 4 & 1 & 8 & 1 \\
            & ZeRO-3 & R & N & R & N & R & N \\
            & MiCS & 8 & 1 & 8 & 1 & 8 & 1 \\
            & ZeRO++\tnote{2} & R & N & R & N & R & N \\ \midrule
            \multirow{5}{*}{LLaMA-30B} & Our Work & 8 & 1 & 8 & 1 & 8 & 4 \\
            & ZeRO-3 & R & N & R & N & R & N \\
            & MiCS & 8 & 2 & 8 & 2 & 8 & 2 \\
            & ZeRO++\tnote{2} & R & N & R & N & R & N \\ \bottomrule
        \end{tabular}
        \begin{tablenotes}
            \footnotesize
            \item[1] We set $s_{dp}^0 = R, s_{dp}^1 = N$ in this experiment.
            \item[2] ZeRO++ uses $s_p^0=8, s_p^1=1$ to shard secondary parameters.
        \end{tablenotes}
    \end{threeparttable}
\end{table}
\renewcommand{\arraystretch}{1}

\section{Evaluation}
\label{sec_evaluation}

\subsection{Experimental Setup}

\subsubsection{Implementation} 

We use an iterative solver to dynamically optimize communication costs based on the provided configuration. 
To uphold comparable computational performance, \SysName incorporates FlashAttention-v2  \cite{dao2023flashattention} and adopts mixed-precision training with BF16, aligning with baseline systems. We also introduce a user-friendly interface enabling users to customize the sharding of $P$, $G$, and $OS$ through predefined configurations or leverage the integrated solver to automatically determine the optimal sharding strategy.

\subsubsection{Testbed} We evaluate the training performance of three popular LLMs: LLaMA-7B, LLaMA-13B, and LLaMA-30B. The training is conducted on a dedicated cluster with 128 GPU servers. Each server is equipped with 8 GPUs and 128 CPU cores, resulting in a total of 1024 NVIDIA Ampere GPUs (A800). Each GPU is outfitted with 80GB of memory, interconnected through NVLink within a node, and inter-node communication is facilitated by 4 Mellanox HDR InfiniBand without SHARP.

\subsubsection{Baselines \& Evaluation Metrics} 

We conduct a comprehensive benchmark of \SysName, comparing it against DeepSpeed-ZeRO1, DeepSpeed-ZeRO3 \cite{ZeRO}, DeepSpeed-ZeRO++\cite{ZeRO++}, and DeepSpeed-MiCS \cite{MiCS}. Our evaluation focuses on key performance metrics, including Model FLOPs Utilization (MFU)\footnote{We calculate FLOPs and MFU using the formula in Megatron-LM. As detailed in \cite{dao2023flashattention}, while the FLOPs due to attention should be halved, with causal mask, only approximately half the number of elements in attention needs computation. For consistency, we adhere to the literature formula (without dividing attention FLOPs by 2) as in FlashAttention and many other libraries.}\cite{palm}   and Tokens per GPU per Second (TGS). The sequence length is held constant at 4096 tokens in all experiments. The sequence length is fixed at 4096 tokens. Micro-batch size is configured to 1 sequence with 4096 tokens, while the global-batch size is set to 4 million tokens. The micro-batch number is 128 with 8 GPUs (i.e. $M=128$); however, training with 1024 GPUs reduces the micro-batch number to 1 (i.e. $M=1$). Since the core objective of \SysName is reducing communication overhead in ZeRO, we adopt $s_{dp}^0 = R, s_{dp}^1 = N$ across all experiments, excluding tensor or pipeline parallelism.

\begin{figure}[t]
    \centering
    \includegraphics[width=\linewidth]{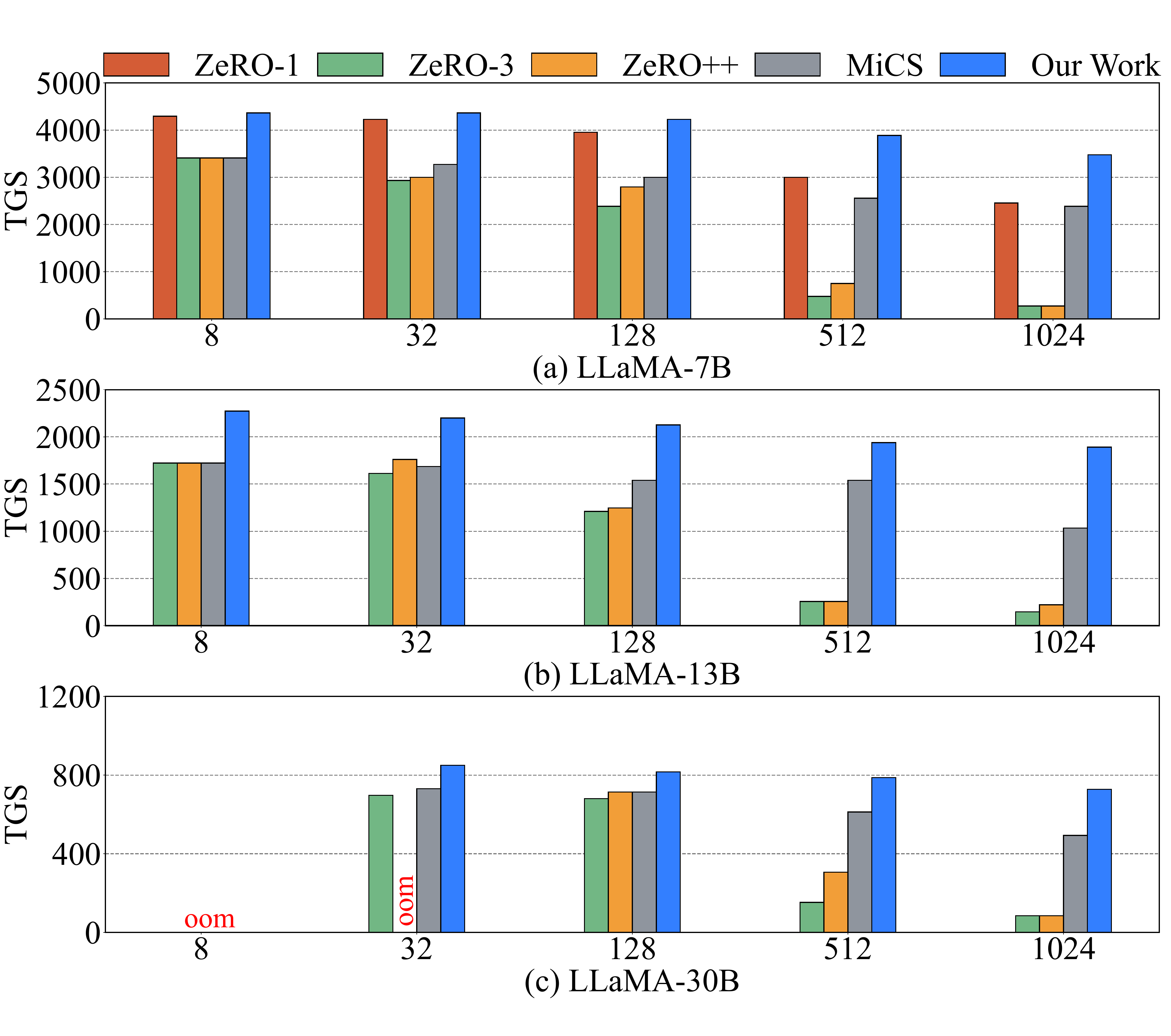}
    \caption{End-to-end evaluation results (TGS) of training LLaMA-based Models from 8 GPUs to 1024 GPUs. }
    \label{e2eTGS}
\end{figure}

\subsubsection{System Configurations}

Table \ref{tab:configs} presents the configurations utilized in \SysName and the baselines. \SysName maintains a uniform set of configurations when scaling training from 8 GPUs to 1024 GPUs. In ZeRO++, the secondary shard number of the parameters is  tuned for optimal performance, with $(s_p^0=8, s_p^1=1)$, and quantization is not enabled. Activation recomputation is applied during the training of LLaMA-30B, while it is disabled for LLaMA-7B and LLaMA-13B. The communication-computation overlap configurations are consistently enabled in all baselines.
To ensure a fair comparison with baselines, we  disabled the overlap between \texttt{Broadcast} and forward computation during the end-to-end system evaluations, as these baselines do not provide this function.

\subsection{End-to-End System Evaluation}
\label{sec:e2e}

% In this section, we present a systematic analysis of the scalability of \SysName from 8, 32, 128, 512 to 1024 GPUs, using LLaMA-7B, LLaMA-13B, and LLaMA-30B.

\subsubsection{Scalability Performance}

Figure \ref{e2eMFU} illustrates the MFU during the training of models of varying sizes with different GPU number, while Figure \ref{e2eTGS} provides corresponding TGS results. \SysName exhibits higher performance across all cases than the basesline systems. Specifically, it achieves 51\%, 52\%, and 42\% MFU when training LLaMA-7B, LLaMA-13B, and LLaMA-30B models with 1024 GPUs, respectively.

When training LLaMA-7B with 8 GPUs, ZeRO-1 exhibits a very similar MFU to \SysName. The observation  indicates that both systems achieve comparable computation efficiency. This similarity arises from the fact that they share the same communication cost. Importantly, this result underscores that \SysName, despite introducing innovative communication optimizations, maintains a comparable level of computation efficiency with baseline systems, given the commonality in utilizing the same computation engine, such as FlashAttention.

% With 8 GPUs, the partitioning in ZeRO-1 aligns with \SysName, while ZeRO3, ZeRO++, and MiCS share a scheme (based on ZeRO3), showing consistent MFUs. When scaling up to less than 512 GPUs, the MFU differences between ZeRO-1 and LinS and between ZeRO3, ZeRO++, and MiCS are minimal. However, expanding to 512 GPUs, these gaps widen, especially for ZeRO3 and ZeRO++, which drop by about 200\%, consistent with Section \ref{subsec:highcommcost} analysis that the communication latency has a tendency to plummet from 128 to 512 GPUs as shown in Figure \ref{fig: motivation}(c).

\begin{figure}[t]
    \centering
    \includegraphics[width=\linewidth]{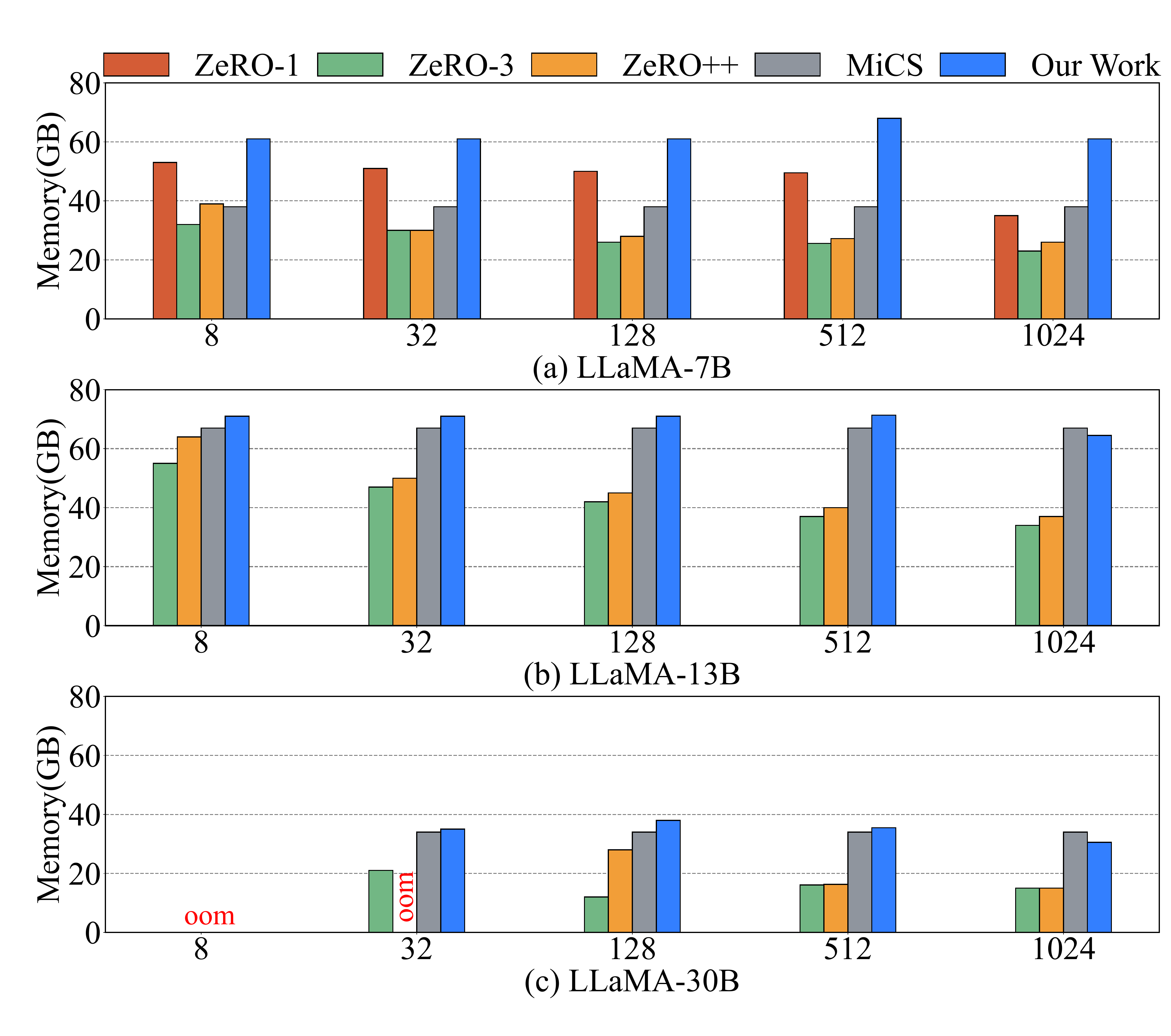}
    \caption{Peak Memory of training LLaMA-7B/13/30B from 8 to 1024 GPUs.}
    \label{e2eMemory}
\end{figure}

As the GPU number increases, \SysName demonstrates a comparatively stable decrease in MFU and TGS across LLaMA-7B, LLaMA-13B and LLaMA-30B models when compared to other baselines. Expanding from 8 to 1024 GPUs, \SysName experiences a modest 15\% reduction in MFU, while ZeRO-3 can exhibit reductions of up to 88\%. The decrease in \SysName's MFU is attributed to the reduced computational load per GPU as the number of GPUs grows, leading to a higher communication-to-compute ratio.
Notably, the MiCS approach, which employs a subgroup communication strategy, also exhibits a relatively gentle downward trend, similar to \SysName. However, due to its limited array of configuration options, MiCS consistently maintains an MFU below that of \SysName.
Zero++, while often outperforming ZeRO-3, faces challenges such as out-of-memory (OOM) issues. For instance, when training LLaMA-30B on 32 GPUs, Zero++ encountered an OOM situation, whereas other methods achieved an MFU of around 40\%.

When training LLaMA-7B on 1024 GPUs, \SysName achieves MFU 51\%, surpassing other baselines. ZeRO-1 follows closely with 36\% MFU, while MiCS ranks third at 35\%. ZeRO-3 and ZeRO++ lag significantly behind, achieving  approximately 4\% MFU.
In comparison to ZeRO-1, \SysName effectively constrains the \texttt{Broadcast} operation, which is used to disseminate updated parameters to other GPUs at the end of each step, involving only 8 GPUs. This strategic approach minimizes communication overhead. On the other hand, MiCS also reduces the communication scale of \texttt{AllGather} and \texttt{ReduceScatter} within a node for parameters fetch and gradients distribution, yet it generates more traffic than \SysName. Consequently, MiCS exhibits lower performer compared to ZeRO-1 when training LLaMA-7B on 1024 GPUs.
For LLaMA-13B and LLaMA-30B training on 1024 GPUs, \SysName maintains its leading position with MFU values of 51\% and 43\%, respectively. MiCS follows with 33\% for LLaMA-13B and 29\% for LLaMA-30B. 
% Compared to MiCS, \SysName adopts configurations that facilitate \texttt{AllGather} communication on fewer GPUs. This  choice enhances communication efficiency while satisfying memory requirements.

% When training LLaMA-7B on 1024 GPUs, \SysName achieves an MFU of 51\%, followed by ZeRO-1 at 36\% MFU, with MiCS in third at 35\% MFU, while ZeRO-3 and ZeRO++ have only approximately 4\% MFU. LinS restricts ZeRO-1's \texttt{AllGather} communication across 1024 GPUs to within 8 GPUs per step, utilizing high-speed intra-node bandwidth for parameter updates and reducing \texttt{AllGather} communication before forward and backward computations compared to MiCS. For LLaMA-13B and LLaMA-30B training on 1024 GPUs, \SysName reaches an MFU of 51\% and 43\%, followed by MiCS of 33\% and 29\%. Compared to MiCS with equivalent settings, ($s_p=s_g=s_{os}=8$ in 13B and $s_p=s_g=s_{os}=16$ in 30B) \SysName adopts configurations that allow for \texttt{AllGather} communication on fewer GPUs, enhancing communication efficiency while meeting memory requirements. 

\begin{figure}[t]
    \centering
    \includegraphics[width=\linewidth]{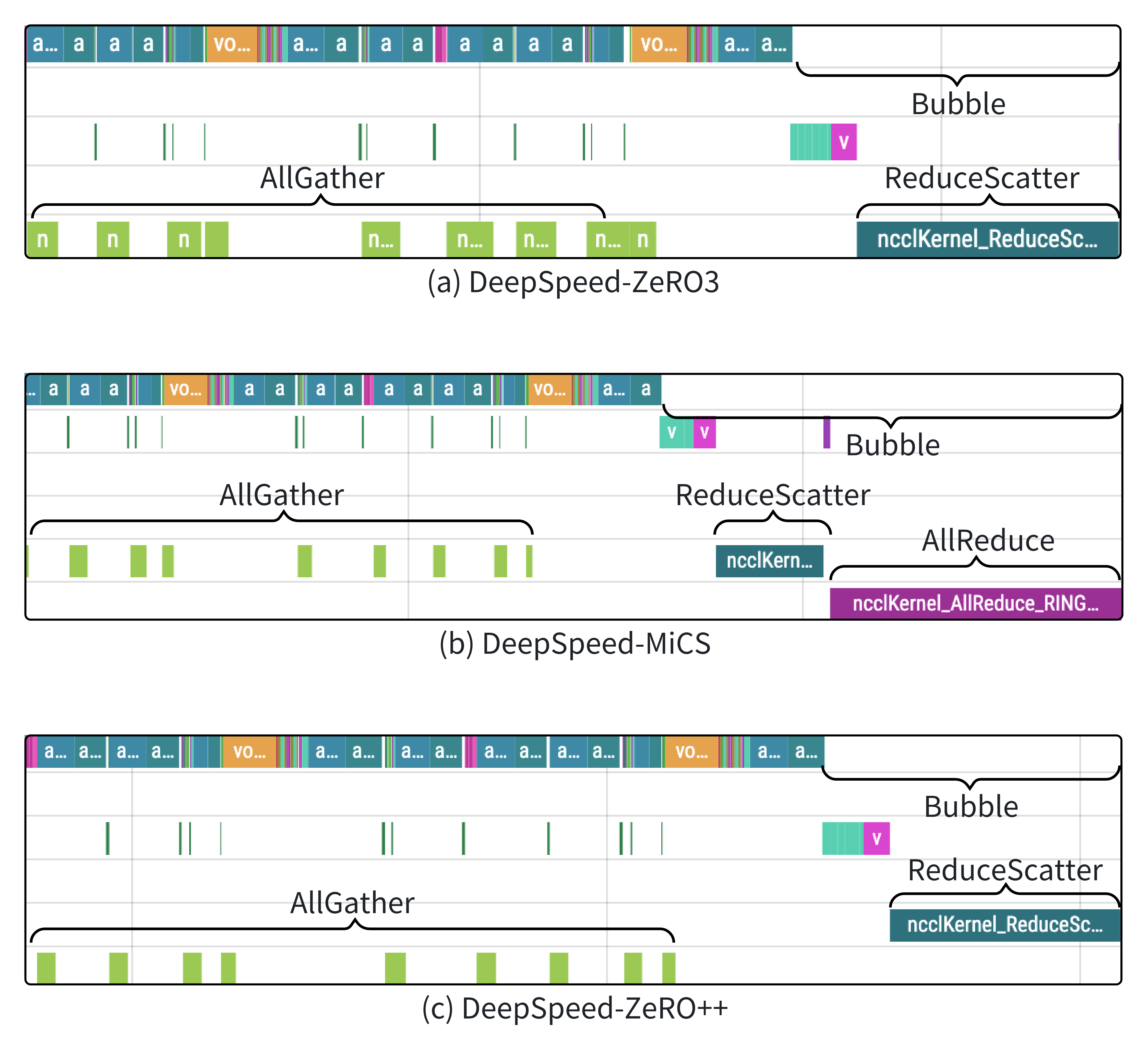}
    \caption{Trace segment for the LLaMA-7B model training on 32 GPUs with a micro-batch size of 4096 tokens using DeepSpeed-ZeRO3/MiCS/ZeRO++. This trace captures the backward phase of the last micro-batch within a step.}
    \label{fig:trace_bubble}
\end{figure}

\begin{figure*}[ht]
    \centering
    \includegraphics[width=\linewidth]{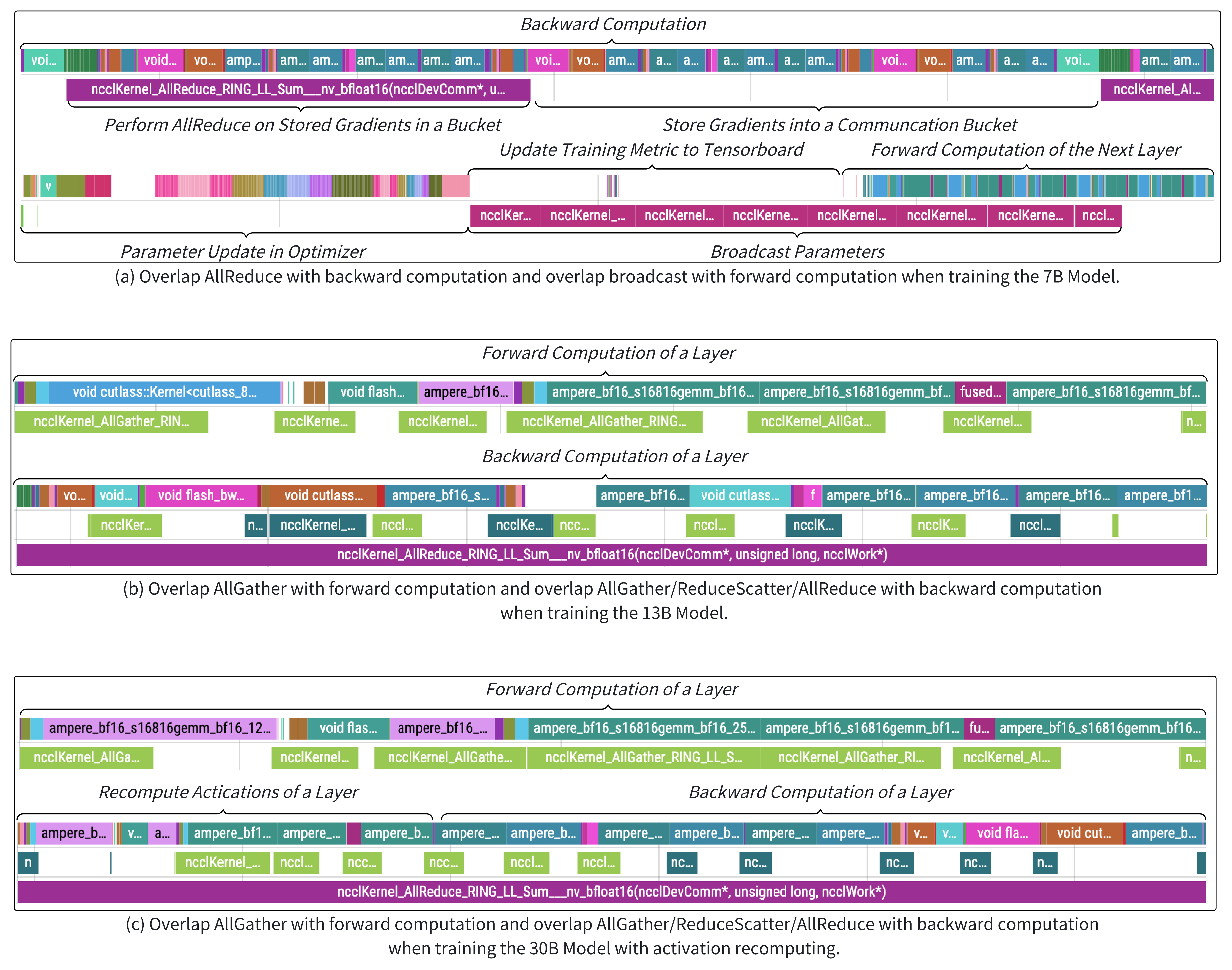}
    \caption{Training trace segment for LLaMA-7B, LLaMA-13B, and LLaMA-30B models using \SysName. This trace encompasses both the forward and backward phases of a training step. LLaMA-7B and LLaMA-13B are trained with a micro-batch size of 4096 tokens and a micro-batch number of 2 on 32 GPUs, while LLaMA-30B utilizes 64 GPUs for training.}
    \label{fig:TraceOverlap}
\end{figure*}

\subsubsection{GPU Memory Analysis}

Figure \ref{e2eMemory} illustrates the maximum allocated memory during training for various systems. ZeRO-3 stands out as the most memory-efficient, primarily due to its aggressive splitting of model states across all GPUs. The memory consumption of ZeRO-3 becomes stable in large-scale training. For instance, when scaling from 512 GPUs to 1024 GPUs, the memory consumption for training LLaMA-30B changes from 16GB to 15GB with ZeRO-3. 
Both ZeRO++ and MiCS demonstrate enhanced training performance at the expense of higher memory consumption. MiCS, in particular, prioritizes redundant storage to optimize communication efficiency, resulting in approximately double the memory usage compared to ZeRO-3 for the same models on 1024 GPUs. This trade-off highlights the strategic use of memory resources to achieve superior training outcomes with these approaches.

\SysName consistently exhibits high memory consumption, particularly noticeable when comparing it to MiCS. For instance, when training the 7B model on 1024 GPUs, \SysName's memory footprint is twice that of MiCS, despite achieving a more efficient utilization of memory. Although \SysName and MiCS demonstrate similar memory consumption for the 13B and 30B models, their memory utilization compositions vary. \SysName attains higher training efficiency by dynamically managing the memory allocation of parameters, gradients and optimizer states. This dynamic allocation strategy allows \SysName to optimize memory usage effectively, contributing to its superior training efficiency despite the higher memory footprint.

\begin{figure}[t]
    \centering
    \includegraphics[width=\linewidth]{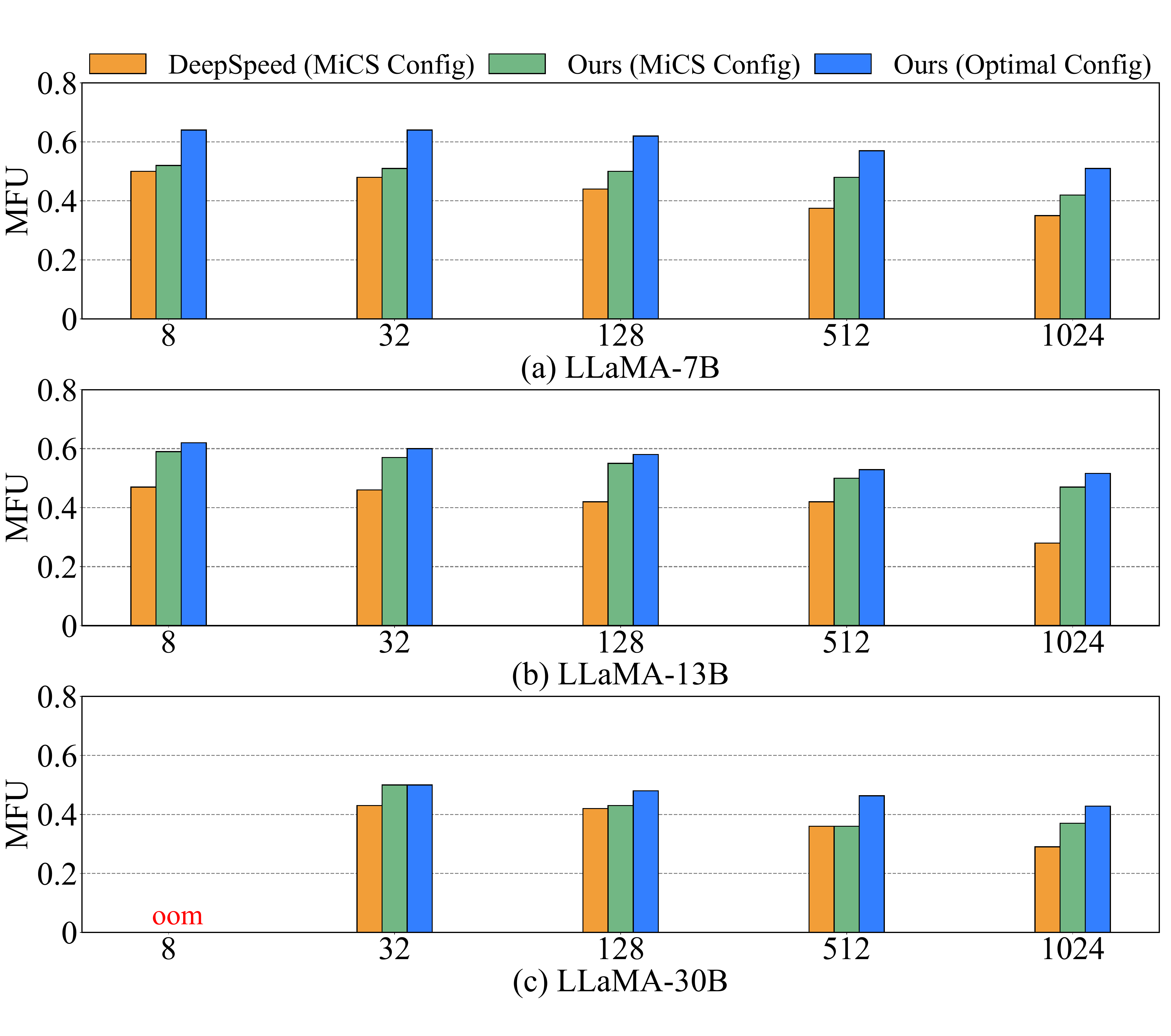}
    \caption{Effects (MFU) of Sharding Strategy in training LLaMA-based Models from 8 GPUs to 1024 GPUs.}
    \label{fig:ABPartition}
\end{figure}

\subsection{Performance Gap Analysis}

Our investigation reveals that the superior training performance of \SysName can be attributed to two crucial optimizations: a flexible sharding strategy for parameters, gradients, and optimizer states, and an advanced methodology for orchestrating the overlap between communication and computation. Our findings underscore that existing implementations of ZeRO-3, MiCS, and ZeRO++ within the DeepSpeed framework encounter significant hurdles in achieving effective overlap between communication and computation, particularly during the backward phases. This challenge is notably evident in the inefficient utilization of idle compute resources during the \texttt{ReduceScatter} communications in ZeRO-3 and ZeRO++, as well as the combined \texttt{ReduceScatter} and \texttt{AllReduce} communications in MiCS, as depicted in Figure \ref{fig:trace_bubble}. In this figure, we present a trace segment for the LLaMA-7B model training on 32 GPUs with a micro-batch size of 4096 tokens using DeepSpeed-ZeRO3/MiCS/ZeRO++. This trace captures the backward phase of the last micro-batch within a step. It reveals instances of computation resource bubbles during the backward pass. Even when the communication-computation overlap setting is enabled, DeepSpeed-ZeRO3/MiCS/ZeRO++ fails to effectively overlap \texttt{ReduceScatter} with computation. Additionally, in DeepSpeed-MiCS, the \texttt{ReduceScatter} operation also blocks the concurrent execution of \texttt{AllReduce}. We shown the complete trace in Appendix \ref{TraceAnalysis}.

\SysName excels in orchestrating the seamless overlap of communications with computation, leading to a substantial improvement in computing resource utilization, as depicted in Figure \ref{fig:TraceOverlap}. During the forward pass, as shown in Figure \ref{fig:TraceOverlap} (b) and (c), \SysName facilitates overlap by concurrently computing each layer alongside the \texttt{AllGather} communications necessary for acquiring the subsequent layer's parameters. During the backward phase, \SysName skillfully overlaps \texttt{AllReduce} and \texttt{ReduceScatter} communications with computation. Moreover, \texttt{AllReduce} can be executed concurrently with \texttt{AllGather} and \texttt{ReduceScatter}. Additionally, \SysName synchronizes the broadcasting of updated parameters with the forward computation of the next step, as illustrated in Figure \ref{fig:TraceOverlap} (a).

\subsection{Ablation Study }

We present ablation experiments to  substantiate the effectiveness of the flexible sharding strategy in \SysName,

\subsubsection{Analysis of Sharding Strategy}

\begin{figure}[t]
    \centering
    \includegraphics[width=\linewidth]{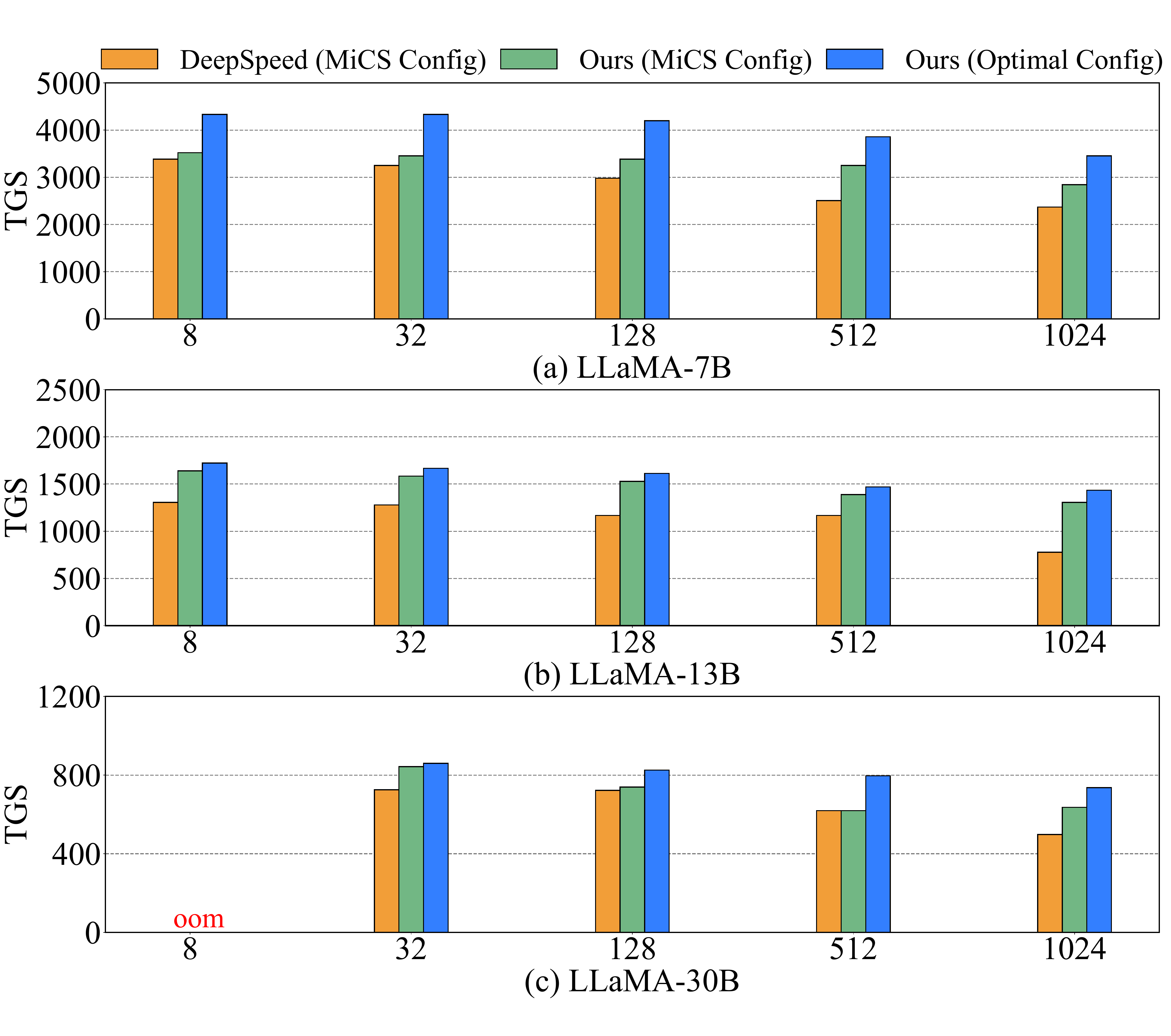}
    \caption{Effects (TGS) of Sharding Strategy in training LLaMA-based Models from 8 GPUs to 1024 GPUs.}
    \label{fig:ABPartitionTGS}
\end{figure}

We conducted experiments to validate the effectiveness of our sharding strategy by comparing our Planner-based optimal configurations with MiCS's rule-based config under the same execution engine. Figure \ref{fig:ABPartition} illustrates the MFU results, while Figure \ref{fig:ABPartitionTGS} provides corresponding TGS results.
Our observations reveal that with a 1024-GPU setup, the optimal configuration for \SysName at model sizes of 7B, 13B, and 30B yields MFU values that are 1.3x, 1.1x, and 1.13x higher, respectively, compared to the MiCS configuration. Moreover, implementing MiCS under \SysName results in higher performance compared to the implementation on DeepSpeed. This improvement can be attributed to our optimizations for communication-computation overlap.

\renewcommand{\arraystretch}{0.95}
\begin{table*}[htbp]
\caption{Effects of Overlap Strategy on Training LLaMA-based Models with Micro-Batch Sizes from 1 to 8 Using 64 GPUs. (TGS and MFU)}
\centering
\label{tab:ABOverlap}
\begin{tabular}{@{}ll|cccc|cccc@{}}
\toprule
& \multirow{2}{*}{\textbf{Overlap Strategy}} & \multicolumn{4}{c}{\textbf{TGS}} & \multicolumn{4}{|c}{\textbf{MFU}}  \\ 
\cmidrule(lr){3-6} \cmidrule(lr){7-10} &  & 
\multicolumn{1}{c}{$M=1$} & \multicolumn{1}{c}{$M=2$} & \multicolumn{1}{c}{$M=4$} & \multicolumn{1}{c}{$M=8$} & 
\multicolumn{1}{|c}{$M=1$} & \multicolumn{1}{c}{$M=2$} & \multicolumn{1}{c}{$M=4$} & \multicolumn{1}{c}{$M=8$}  \\ \midrule
\multirow{4}{*}{7B}    & Overlap AllGather/ReduceScatter+AllReduce+Broadcast  & 3525  & 4119 & 4416  & 4503 & 0.57 & 0.62 & 0.65 & 0.67  \\
& Overlap AllGather/ReduceScatter+AllReduce  & 3346 & 3991 & 4329 & 4438 & 0.54  & 0.60 & 0.64  & 0.65 \\
& Overlap AllGather/ReduceScatter   & 2812   & 3609  & 4070  & 4286  & 0.45  &0.54   & 0.60&  0.63 \\
& No Overlap   & 2812   & 3609  & 4070  & 4286  & 0.45  &0.54   & 0.60&  0.63 \\ \midrule
\multirow{4}{*}{13B}   & Overlap AllGather/ReduceScatter+AllReduce+Broadcast  & 1918  & 2082 & 2183 & 2230 & 0.53 & 0.58  & 0.61 & 0.62  \\
& Overlap AllGather/ReduceScatter+AllReduce  & 1895 & 1920 & 2160 & 2184 &  0.53 & 0.54 & 0.60 & 0.61 \\
& Overlap AllGather/ReduceScatter   & 1630   & 1782  &  2033 &  2126 & 0.45  & 0.49  & 0.56 & 0.59  \\
& No Overlap   &  1527 & 1552 &  1608 &  1666 &  0.42 & 0.43  & 0.44 & 0.46 \\ \midrule
\multirow{4}{*}{30B}   & Overlap AllGather/ReduceScatter+AllReduce+Broadcast  & 825 & 856 & 872 & 880 & 0.48 & 0.50 & 0.50 & 0.51 \\
& Overlap AllGather/ReduceScatter+AllReduce  & 759 & 814 & 847 & 859 & 0.44  & 0.47 & 0.49 & 0.50 \\
& Overlap AllGather/ReduceScatter   & 651   & 740  &  814 &  842 & 0.37  & 0.43  & 0.47 & 0.49  \\
& No Overlap  &  557  & 654  & 678  & 719  & 0.33  & 0.38  & 0.39 & 0.42   \\ \bottomrule
\end{tabular}
\end{table*}
\renewcommand{\arraystretch}{1}

\subsubsection{Analysis of Overlap}
% \begin{figure}[t]
%     \centering
%     \includegraphics[width=\linewidth]{ABOverlap.pdf}
%     \caption{Effects of Overlap Strategy in training LLaMA-based Models from micro batch number of 1 to 8 with 64 GPUs is used. \todo{if there is no obvious difference, use table}}
%     \label{fig:ABOverlap}
% \end{figure}

% \begin{table}[]
% \caption{Effects (MFU) of Overlap Strategy in training LLaMA-based Models from micro batch number of 1 to 8 with 64 GPUs is used.}
%     \centering
% \label{tab:ABOverlap}
% \begin{tabular}{@{}lcccc@{}}
% \toprule
% 7B & 1 & 2 & 4 & 8 \\ \midrule
% Full Overlap & 0.57 & 0.62 & 0.65 & 0.67 \\
% Overlap w/o Broadcast & 0.54 & 0.60 & 0.64 & 0.66 \\
% Overlaps w/o Broadcast \& AllReduce & 0.46 & 0.54 & 0.60 & 0.63 \\
% No Overlap & 0.46 & 0.54 & 0.60 & 0.63 \\ \midrule
% 13B & 1 & 2 & 4 & 8 \\ \midrule
% Full Overlap & 0.54 & 0.58 & 0.61 & 0.62 \\
% Overlap w/o Broadcast & 0.53 & 0.54 & 0.60 & 0.61 \\
% Overlaps w/o Broadcast \& AllReduce  & 0.46 & 0.5 & 0.57 & 0.59 \\
% No Overlap & 0.43 & 0.43 & 0.45 & 0.46 \\ \midrule
% 30B & 1 & 2 & 4 & 8 \\ \midrule
% Full Overlap & 0.48 & 0.5 & 0.51 & 0.51 \\
% Overlap w/o Broadcast & 0.44 & 0.48 & 0.5 & 0.5 \\
% Overlaps w/o Broadcast \& AllReduce  & 0.38 & 0.43 & 0.57 & 0.58 \\
% No Overlap & 0.34 & 0.38 & 0.4 & 0.42 \\ \bottomrule
% \end{tabular}
% \end{table}

We conducted a series of experiments to assess the efficacy of our approach in communication-computation overlap. Our investigation involved systematically deactivating overlap optimizations in the following sequence: initially, we disabled the overlap of \texttt{Broadcast} operations for spreading updated parameters. Subsequently, we turned off the overlap of \texttt{AllReduce} for gradient synchronization. Finally, we deactivated the overlap of \texttt{AllGather} and \texttt{ReduceScatter} in parameter sharding, resulting in a state of no overlap during the training process. 
The experiments were conducted using 64 GPUs, and we varied the computational loads by adjusting the micro-batch number $M$. The results of these experiments, showcasing the effects of overlap, are presented in Table \ref{tab:ABOverlap}.

% Generally, as $M$ increased, the overall MFU also rose; however, the relative benefit of overlap diminished due to the larger computational volume leading to an increased computation-to-communication ratio. The results, depicted in Table \ref{tab:ABOverlap}, indicate a stepwise decline in overall MFU as we sequentially deactivated overlap optimizations, highlighting the significant influence of our overlap optimization on performance.

% To validate the effectiveness of our overlap between computation and communication in parameter sharding, as well as the efficiency of overlapping \texttt{AllReduce} and \texttt{Broadcast} operations with computation in optimizer states sharding, we conducted experiments. We systematically disabled overlap optimizations in the following sequence: first, we turned off the overlap of \texttt{Broadcast} in optimizer states sharding; next, we disabled the overlap of \texttt{AllReduce} gradient communication in optimizer states sharding. Finally, we deactivated the overlap of \texttt{AllGather} and \texttt{ReduceScatter} communications in parameter sharding, achieving a state of no overlap during training.

In large-scale distributed LLM training, particularly with 1024 GPUs, the limitations imposed by the global batch size necessitate setting the micro-batch number $M$ to 1. Under such conditions, the impact of our overlap optimizations becomes significantly more pronounced. For instance, in our experiments with a 7B model, disabling the overlap of \texttt{Broadcast} leads to a 5\% drop in MFU. Further disabling the overlap optimization for \texttt{AllReduce} causes an additional 16\% decline in MFU.
When investigating scenarios with $s_p>1$ and $s_{os}>1$, overlaps for \texttt{Broadcast}, \texttt{AllReduce}, \texttt{AllGather}, and \texttt{ReduceScatter} significantly contribute to MFU improvements. For the 13B model, communication overlap could enhance the overall system performance by a factor of 1.25 compared to the case without any overlap setting. Meanwhile, for the 30B model, the improvement is even more substantial, reaching a factor of 1.48. These values  demonstrate the essential nature of overlap optimizations in boosting the efficiency of LLM training.

\section{Related Work}
\noindent\textbf{Model parallelism and 3D parallelism.} Model parallelism is represented by two approaches: tensor parallelism and pipeline parallelism. Tensor parallelism~\cite{Megatron-LM} involves partitioning specific layer weights and introducing additional AllReduce communication. Pipeline parallelism\cite{GPipe,DAPPLE,PipeDream,PipeMare} divides the layers of the model horizontally among each rank. Recent innovations have proposed methods that autonomously discern parallelism approaches by intricately melding both data and model parallelism for distinct operators within the model. To illustrate, solutions like Alpa \cite{Alpa}, OptCNN \cite{OptCNN}, FlexFlow \cite{FlexFlow,Unity}, and TensorOpt \cite{TensorOpt} incorporate both data and tensor parallelism. These leverage a variety of search algorithms to refine and enhance the execution of blueprints. However, while these automated parallelism solutions focus on optimizing the sharding and placement strategies for the optimal operators within the computational graph, they overlook strategies related to the orthogonal placement of the model states.

\noindent\textbf{Large-scale communication optimization.} 
% With the increase in training scale, the improvement in training efficiency achieved through greater computational investment gradually diminishes. 
Some works\cite{ByteScheduler, P3, PyTorchFSDP, sun2019gradientflow} try to overlap communication with computation to mitigate communication costs. 
ZeRO++ and Espresso\cite{Hi-SpeedDNNTrainingwithEspresso} utilize quantization and compression techniques to reduce communication volume, albeit at the expense of precision. 
DEAR\cite{DeAR} aggregates multiple small communications using fixed-size buffers to reduce communication overheads. Hetu\cite{HetuMoE} leverages hierarchical all-to-all to minimize inter-node communication volume under poor inter-node communication. 
% Recognizing that the performance of collective communication primitives significantly degrades with increasing training scope, 
Similarly, Hybrid AllReduce\cite{HighlyScalable} attempts to decompose a single collective communication primitive into a combination of multiple subgroup communications.

\section{Conclusion}

We propose  \SysName to address the communication challenge of distributed  LLM training at scale with ZeRO. 
The proposed \SysName introduces a novel approach by incorporating flexible sharding strategies—\emph{Full-Replica}, \emph{Full-Sharding}, and \emph{Partial-Sharding}—for each component within the model states (Parameters, Gradients, and Optimizer States).
The introduced sharding factors ($s_p^0 \times s_p^1, s_g^0 \times s_g^1, s_{os}^0 \times s_{os}^1$) control GPU and device mesh sharding. Analyzing memory and communication costs for each dimension, \SysName formulates an optimization problem to find factors optimizing communication costs under memory constraints. Additionally, it implements an execution engine tailored for LLM training, and the customized communication and computation overlap strategy, incorporating these flexible sharding factors to achieve optimized communication efficiency during training.  Compared to MiCS and ZeRO++, \SysName improves the training throughput by $1.4-12.7$. 

% \balance

% focusing on the memory and communication overhead of existing strategies, particularly ZeRO, ZeRO++, and MiCS.  Through extensive evaluations on LLaMA-based models, LinS demonstrates superior performance, achieving higher Model FLOPs Utilization (MFU) compared to ZeRO++ and MiCS across various model sizes and GPU configurations.

% Large Language Models (LLMs) are increasingly being trained with more tokens but smaller model sizes. The traditional Zero Redundancy Optimizer (ZeRO) struggles to adapt to this new trend. To address this, we introduced \SysName, a novel training framework. This framework efficiently partitions model states and optimally manages data placement, achieving a 90\% scaling efficiency on 1024 GPUs.

% \todo{It is hard to have a crystal ball, but it could be that small models distilled from larger ones have better performance than training small models directly. This is a minor nit bur was not addressed in the paper.}

% \todo{There was no mention of stragglers. This left me wondering .. there are different ways of dealing with this .. some could compromise convergence and others would not. Would be useful to have some more discussion of this issue.}

% \begingroup
% \tiny
\bibliographystyle{IEEEtran}
\bibliography{references.bib}
% \endgroup

% \clearpage

% \newpage

\begin{figure*}[ht]
    \centering
    % \rotatebox[origin=c]{90}{
        \includegraphics[width=0.95\linewidth]{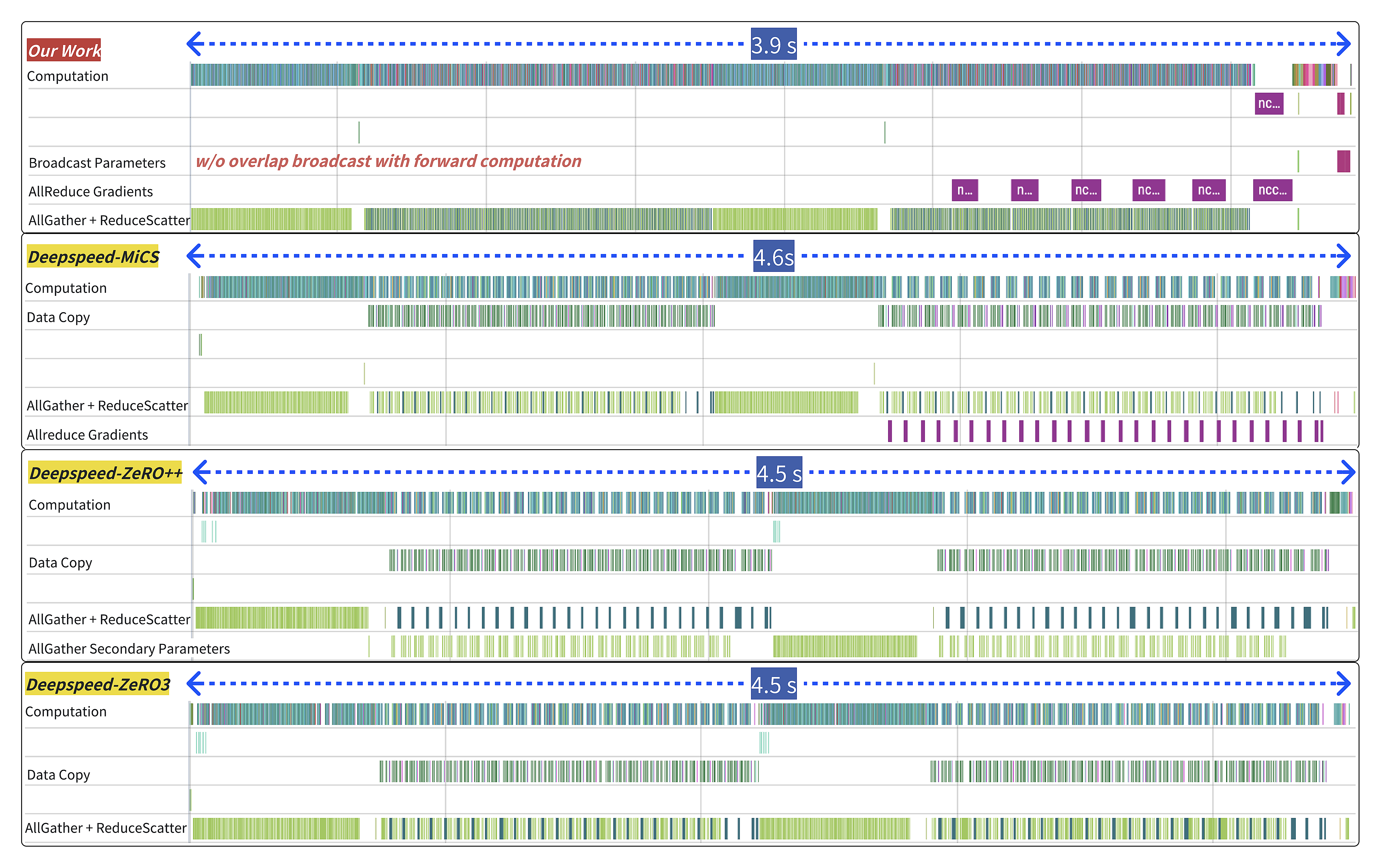}
    % }
    \caption{Training trace of LLaMA-7B on 32 NVIDIA A800 GPUs with a micro-batch size of 4096 tokens and micro-batch number of 2.}
    \label{trace_7b}
\end{figure*}

\appendix

\subsection{Trace Analysis (7B)}
\label{TraceAnalysis}

\begin{figure*}[ht]
    \centering
    % \rotatebox[origin=c]{90}{
        \includegraphics[width=0.95\linewidth]{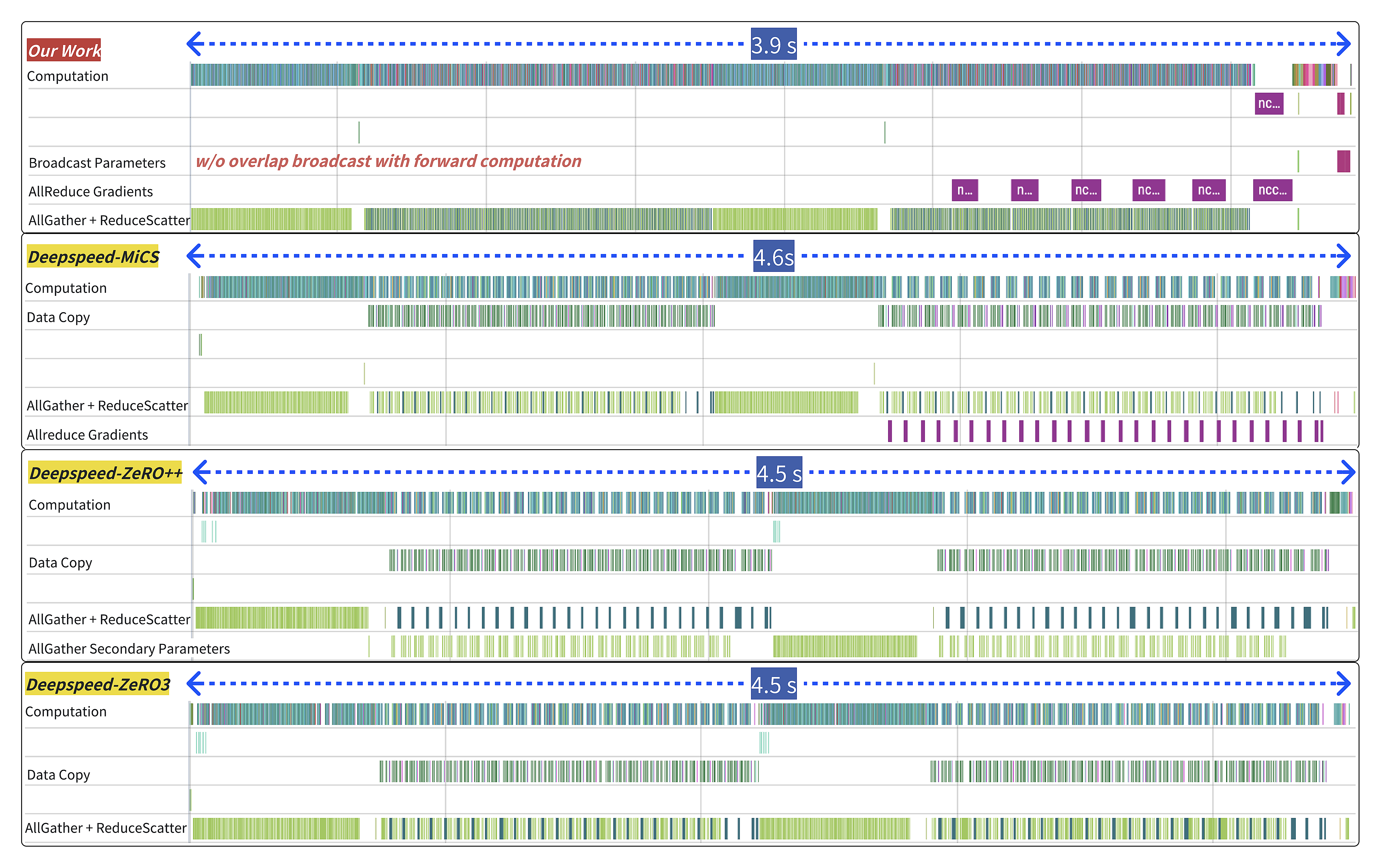}
    % }
    \caption{Training trace of LLaMA-13B on 32 NVIDIA A800 GPUs with a micro-batch size of 4096 tokens and micro-batch number of 2.}
    \label{trace_13b}
\end{figure*}

\begin{figure*}[ht]
    \centering
    % \rotatebox[origin=c]{90}{
        \includegraphics[width=0.95\linewidth]{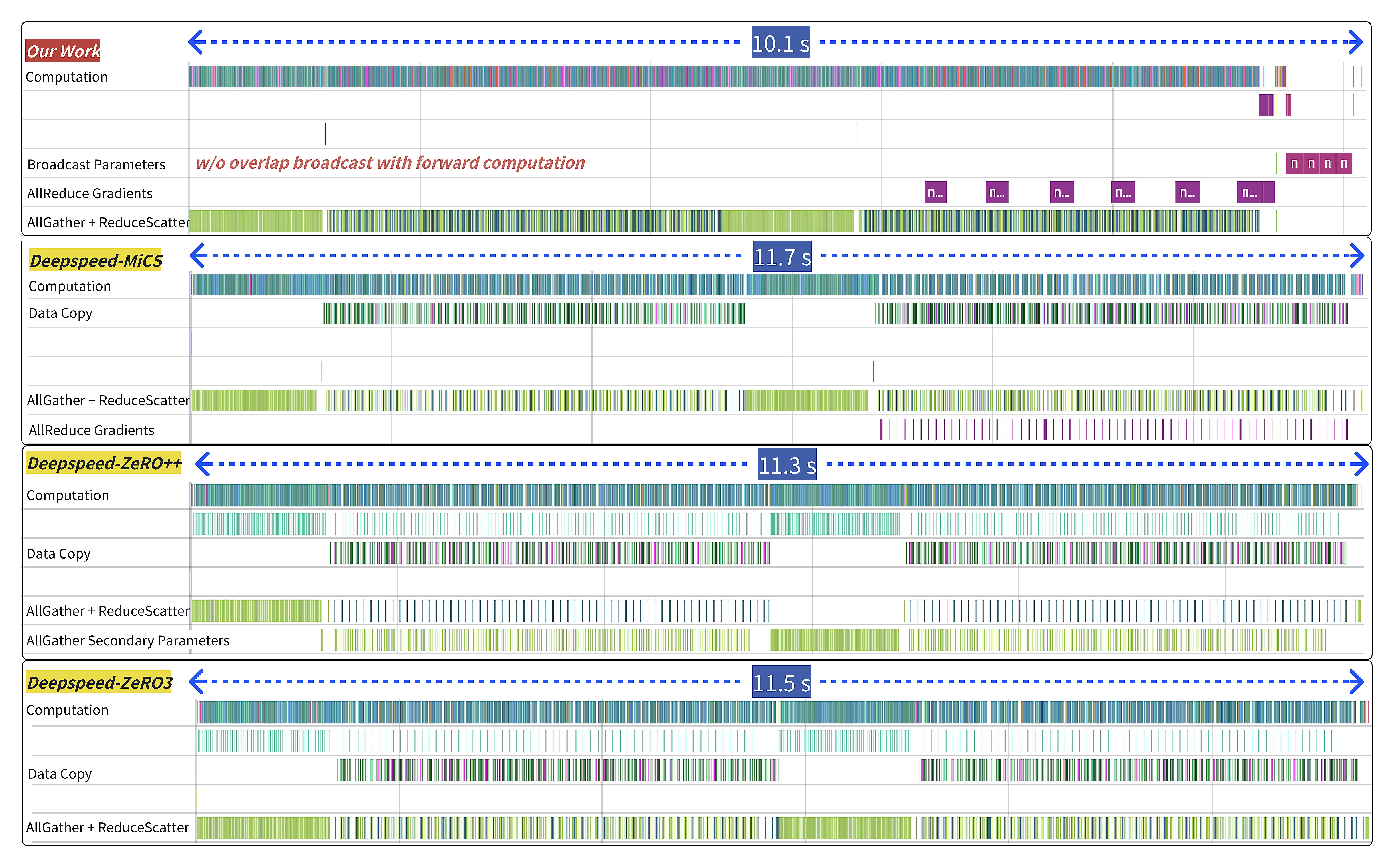}
    % }
    \caption{Training trace of LLaMA-30B on 64 NVIDIA A800 GPUs with a micro-batch size of 4096 tokens and micro-batch number of 2.}
    \label{trace_30b}
\end{figure*}

We present the training traces during a single step for the LLaMA-7B, LLaMA-13B, and LLaMA-30B models under the DeepSpeed framework configurations of ZeRO-3, MiCS, and ZeRO++, as well as under the \SysName framework with its optimal configuration, in Figure \ref{trace_7b}, Figure \ref{trace_13b}, and Figure \ref{trace_30b}, respectively. For this analysis, we utilized a micro batch size of 4096 tokens and a micro batch number of 2 across 32 GPUs for both the 7B and 13B models, while the training of the 30B model was conducted on 64 GPUs. Each GPU is equipped with 80GB of memory, interconnected through NVLink within a node. Inter-node communication is facilitated by 4 Mellanox HDR InfiniBand connections without SHARP.

\end{document}